\documentclass{SciPost}

\hypersetup{
    colorlinks,
    linkcolor={red!50!black},
    citecolor={blue!50!black},
    urlcolor={blue!80!black}
}

\usepackage[bitstream-charter]{mathdesign}
\DeclareSymbolFont{usualmathcal}{OMS}{cmsy}{m}{n}
\DeclareSymbolFontAlphabet{\mathcal}{usualmathcal}

\fancypagestyle{SPstyle}{
\fancyhf{}
\lhead{\colorbox{scipostblue}{\bf \color{white} ~SciPost Physics }}
\rhead{{\bf \color{scipostdeepblue} ~Submission }}

\fancyfoot[C]{\textbf{\thepage}}
}

\usepackage{bbm}
\usepackage{amsmath}
\numberwithin{equation}{section}

\newcommand{\Tr}{\mbox{Tr}\,}
\newcommand{\e}{\mathrm{e}}
\newcommand{\im}{\mathrm{i}}

\begin{document}

\pagestyle{SPstyle}

\begin{center}{\Large \textbf{\color{scipostdeepblue}{
Bulk-boundary correspondence in topological two-dimensional non-Hermitian systems: Toeplitz operators and singular values
}}}\end{center}

\begin{center}\textbf{
Jesko Sirker
}\end{center}

\begin{center}
Department of Physics and Astronomy, University of Manitoba, Winnipeg, Canada R3T 2N2
\\[\baselineskip]
\href{mailto:sirker@physics.umanitoba.ca}{\small sirker@physics.umanitoba.ca}
\end{center}

\section*{\color{scipostdeepblue}{Abstract}}
\textbf{\boldmath{
In contrast to eigenvalue-based approaches, we formulate the bulk–boundary correspondence for two-dimensional non-Hermitian quadratic lattice Hamiltonians in terms of Toeplitz operators and singular values, which correctly capture the stability, localization, and scaling of edge and corner modes. We show that singular values, rather than eigenvalues, provide the only stable foundation for topological protection in non-Hermitian systems because they remain robust under translational-symmetry-breaking perturbations that destabilize the eigenvalue spectrum, rendering it unsuitable for topological classification. Building on Toeplitz operator theory, we establish general results for non-Hermitian Hamiltonians defined on half and quarter planes, relating the topological indices of the associated Toeplitz operators to the number of finite-size singular values that are separated from the bulk singular-value spectrum and vanish in the thermodynamic limit. This yields a precise bulk–boundary correspondence for edge and corner modes, including higher-order topological phases, without requiring crystalline symmetries. We illustrate our general results with detailed examples exhibiting topologically protected families of edge states, coexisting edge and corner modes, and phases with both gapped bulk and edges supporting only stable corner modes. The latter is exemplified by a non-Hermitian generalization of the Benalcazar–Bernevig–Hughes model.
}}

\vspace{\baselineskip}

%\linenumbers

\vspace{10pt}
\noindent\rule{\textwidth}{1pt}
\tableofcontents
\noindent\rule{\textwidth}{1pt}
\vspace{10pt}

%%%%%%%%%% BLOCK: Copyright information
% This block will be filled during the proof stage, and finilized just before publication.
% It exists here only as a placeholder, and should not be modified by authors.
% \noindent\textcolor{white!90!black}{%
% \fbox{\parbox{0.975\linewidth}{%
% \textcolor{white!40!black}{\begin{tabular}{lr}%
%   \begin{minipage}{0.6\textwidth}%
%     {\small Copyright attribution to authors. \newline
%     This work is a submission to SciPost Physics. \newline
%     License information to appear upon publication. \newline
%     Publication information to appear upon publication.}
%   \end{minipage} & \begin{minipage}{0.4\textwidth}
%     {\small Received Date \newline Accepted Date \newline Published Date}%
%   \end{minipage}
% \end{tabular}}
% }}
% }

\section{Introduction}
Ever since the discovery of the quantum Hall effect \cite{vonKlitzing} and the realization that the quantization of the Hall conductance is topological in origin with integers given by Chern numbers \cite{TKNN}, topology has played a prominent role in condensed matter physics and quantum science more generally. For Hermitian Gaussian fermionic systems, topological order has been classified with respect to parity, time reversal, and chiral symmetry leading to the tenfold way \cite{RyuSchnyder,RyuSchnyderReview}. A crucial point is that topology and a topological classification cannot rely on translational invariance. To cite from Ryu {\it et~al.} \cite{RyuSchnyder}: "Thus, the topological features must be properties of a translationally not invariant Hamiltonian. Seeking a classification scheme, one must have some framework within which to classify such translationally not invariant Hamiltonians." In particular, this means that topological properties must remain if translational invariance is broken as long as any relevant symmetry protecting this order---if any---remains intact. For Hermitian systems, one of the most profound consequences of non-trivial topological order is the bulk-boundary correspondence \cite{Hatsugai1993PRL,Kellendonk2002RMP,Graf2013CMP,Prodan2016Book}. If the bulk of a system is topologically non-trivial, as signaled by a non-zero topological index, then at the boundary of this system with a topologically trivial one or a system with a topologically different index, there will be gapless modes. For a topological insulator, in particular, this means that the boundaries support gapless excitations while the bulk is gapped. In two and higher dimensions, higher-order topology is also possible where only the boundaries of boundaries (corners in two dimensions and hinges in three dimensions) support gapless excitations \cite{Benalcazar2017Science,Langbehn2017PRL,Song2017PRL,Schindler2018NatPhys}. Higher-order topology was first explained using additional crystalline order \cite{Trifunovic2019PRX} but later has been shown to persist even if these crystalline symmetries are broken \cite{Hayashi2019Toeplitz}. Another interesting aspect of topological insulators is that their ground state always has non-trivial entanglement which, in the case of particle conservation, can be further split into number and configurational entropy. For both entropies, non-trivial bounds exist if the system is non-trivial \cite{MonkmanSirker2,MonkmanSirker3,MonkmanSirker4}. 

From a mathematical perspective, a bulk–boundary correspondence is fundamentally a statement about operators on infinite lattices and their truncations to finite domains. Topological indices characterize stability properties of these operators under local perturbations, while boundary and corner phenomena arise from the interplay between bulk topology and the geometry of the truncation. Translational invariance and Bloch Hamiltonians provide a convenient representation of bulk operators, but the existence and robustness of boundary phenomena are ultimately operator-theoretic properties that do not rely on momentum-space descriptions.

In non-Hermitian systems, a comprehensive bulk classification based on point and line gaps has been established \cite{BernardLeClair2002NonHermitianRandomMatrices,Gong2018PRX,Okuma2019PRL,Shiozaki2019PRB}. These works provide a homotopy classification of Bloch Hamiltonians in the absence of boundaries. However, the relation between bulk topological classifications and what happens in finite systems with boundaries is much more complex and still not fully resolved. The main issue is that in non-Hermitian systems the eigenspectrum is generically unstable in contrast to the Hermitian case. A change of boundary conditions or a breaking of translational invariance can change the entire spectrum and there is also no uniform convergence of the eigenvalues with system size. From this perspective, the large pseudospectra typical of non-Hermitian operators reflect the instability of eigenvalues rather than any physical boundary phenomena. As a consequence, attempts to apply an eigenvalue-based bulk-boundary correspondence to non-Hermitian systems fail. Approaches based on non-Bloch band theory or generalized Brillouin zones remain eigenvalue-based and therefore inherit the intrinsic spectral instability of non-normal operators discussed further in Sec.~\ref{Sec_EV}.

From an operator theory perspective, this failure does not come as a surprise: index theorems and K-splitting theorems generally connect bulk topological indices to properties of the {\it singular value spectrum}, rather than to eigenvalues \cite{BoettcherSilbermann}. For Hermitian systems this distinction does not matter because in this case the singular values are just the absolute values of the eigenvalues. However, this is no longer the case for non-Hermitian systems where the singular value and the eigenvalue spectrum behave fundamentally differently \cite{TrefethenEmbree2005}. While the singular value spectrum is mathematically the correct quantity in topological non-Hermitian systems, it has so far only been sparsely considered in the physics literature \cite{HerviouBardarson,BrunelliWanjura,Porras1,Porras2}. Very recently, a proper bulk-boundary correspondence was established for one-dimensional non-Hermitian systems based on Toeplitz operator theory and K-splitting theorems for the singular value spectrum \cite{MonkmanSirker_nH}. Here it was shown, in particular, that a system which has an exact zero-energy edge mode in the semi-infinite limit does not have to have an exact edge mode with exponentially small energy in a finite system; a concept we are used to in the Hermitian case. Instead, the finite system only has an extremely long-lived approximate eigenmode which is mapped exponentially close to zero by the finite-system Hamiltonian. These modes correspond to exponentially small singular values which are separated from the bulk singular values by a spectral gap. The number of such modes is directly related to the topological index.

The natural mathematical framework for addressing these questions is provided by Toeplitz operator theory \cite{BoettcherSilbermann,BoettcherSilbermann2006,Hayashi2021Classification,Hayashi2023QuarterPlane}. A translation-invariant lattice Hamiltonian defines an operator on an infinite Hilbert space whose matrix elements depend only on relative coordinates. Boundaries and corners correspond to truncations of this operator to half-spaces or quarter-spaces, which generically destroy translational invariance but preserve the underlying Toeplitz structure. Classical index theorems characterize the kernel of the corresponding infinite Toeplitz operator, while K-splitting theorems explain how these infinite-volume zero modes manifest themselves in finite systems not as exact eigenstates, but rather as exponentially small singular values separated from the bulk by a spectral gap, which remains open in the thermodynamic limit. In this sense, Toeplitz theory provides the precise mathematical formulation of bulk–boundary correspondence in non-Hermitian systems.

In this work, we will lay out in some detail the fundamentals of Toeplitz theory and the related index theorems as far as they are relevant for the physical properties of topological non-Hermitian systems. We will then develop a bulk–boundary correspondence for two-dimensional non-Hermitian lattice Hamiltonians based on the singular value spectrum of finite truncations of these Toeplitz operators. Our approach builds on classical results in multi-dimensional Toeplitz operator theory \cite{BoettcherSilbermann} and avoids reliance on eigenvalue spectra, which are generally unstable in non-Hermitian systems. It is also important to note that in two dimensions, different boundary geometries (edges versus corners) correspond to fundamentally different truncations of the underlying Toeplitz operators and require distinct index-theoretic treatments. As in the one-dimensional case \cite{MonkmanSirker_nH}, the index in general controls only the minimal number of exponentially small singular values, not their exact multiplicity.

Our work is organized as follows: Sec.~\ref{Sec_Fun} recalls some fundamental results about the eigenvalue and singular value spectra of non-Hermitian Hamiltonians and also introduces Toeplitz operators, index theorems for such operators, and discusses finite truncations. Finally, we also recall aspects of polar and singular value decompositions which are useful in this context. Sec.~\ref{Sec_gen} presents our general results for the bulk-boundary correspondence in two dimensions, both for scalar and for matrix-valued symbols. Sec.~\ref{Sec_HN} discusses as a first example a two-dimensional Hatano-Nelson model on the square lattice. We show that this model has families of edge modes. Sec.~\ref{Sec_HN_ext} extends the Hatano-Nelson model so that a phase with non-zero winding numbers in both spatial directions does exist but we show that this is not sufficient to support a stable corner mode. In Sec.~\ref{Sec_cor}, we then construct a model where, in addition to a family of edge modes, a spectrally protected stable corner mode exists and explain why this is the case. Sec.~\ref{Sec_BBH} introduces a non-Hermitian generalization of the Benalcazar-Bernevig-Hughes model and derives a bulk-boundary correspondence between winding numbers and protected singular values and their corresponding topologically protected corner modes. In Sec.~\ref{Sec_Con}, we summarize our main results.

\section{Fundamentals}
\label{Sec_Fun}
The main purposes of this section are, on the one hand, to recall some basic theorems which explain why bulk-boundary correspondences built on eigenspectra cannot succeed in non-Hermitian systems and, on the other hand, to establish the basics of Toeplitz operator theory including index theorems and their relation to singular values and K-splitting theorems. We will also briefly recall aspects of singular value and polar decompositions which are useful in this context.

\subsection{Connection to open systems}
To set up the problem, we begin by discussing why non-Hermitian quantum systems are important and how they naturally arise, for example, in the study of open quantum systems when quantum jumps can be ignored. A natural starting point is the Lindblad Master equation for the density matrix $\rho$ of the system
\begin{equation}
    \label{Lindblad}
    \dot\rho =-\im[H,\rho] + \sum_\alpha \left(L_\alpha \rho L_\alpha^\dagger -\frac{1}{2}\{L_\alpha^\dagger L_\alpha,\rho\}\right) \, .
\end{equation}
Here $H$ is the Hamiltonian of the system and $L_\alpha$ are dissipators describing the coupling of the system to the environment. The curly bracket denotes the anti-commutator. We are setting $\hbar\equiv 1$ throughout the paper. The first term represents the standard von-Neumann term describing the unitary evolution of the system while the second term describes stochastic quantum jump processes associated with the coupling of the system to the environment. The third term can be understood as a non-unitary damping. If one is interested in the evolution of the system without quantum jumps---either because one can parametrically justify ignoring them in the time regime of interest or one performs a postselection onto quantum trajectories without jumps---then one can define an effective non-Hermitian Hamiltonian by
\begin{equation}
    \label{Heff}
    H_{\rm eff} = H -\frac{\im}{2}\sum_\alpha L_\alpha^\dagger L_\alpha \, .
\end{equation}
It is straightforward to check that $H_{\rm eff}$ reproduces the Lindblad equation \eqref{Lindblad} without the quantum jump term if it satisfies the generalized von-Neumann equation $\dot\rho = -i (H_{\rm eff}\rho -\rho H_{\rm eff}^\dagger)$. Since $H_{\rm eff}$ is generically non-normal (i.e.~it does not commute with its adjoint), its spectral properties differ qualitatively from those of Hermitian operators, a point that will play a central role in the following. Another important point to note is that $L^\dagger_\alpha L_\alpha$ is a Hermitian positive semidefinite operator. As a consequence, an effective non-Hermitian Hamiltonian coming purely from a Lindblad equation will always contain a loss term. However, by first considering a system with additional ancilla sites and then projecting onto the subsystem under investigation, one can also generate an effective gain term which cancels this loss. We do not want to go into these details much further here but these considerations already clearly show that non-Hermitian systems are not just a mathematical curiosity but that there is a strong physical motivation to study them. While the above open-system perspective provides a natural physical motivation to study non-Hermitian Hamiltonians, we will treat general bounded non-Hermitian lattice operators here without assuming an underlying Lindblad structure. Our goal is to derive results which are as general as possible without making assumptions about the origin of the non-Hermitian operators. In the following, we will be concerned with general non-Hermitian lattice Hamiltonians acting on infinite or semi-infinite lattices, whose translationally invariant bulk structure leads naturally to Toeplitz operators.

\subsection{Instability of the eigenspectrum}
\label{Sec_EV}
For a Hermitian system, we are used to considering only the eigenvalues when analyzing its spectral properties. In fact, we are so accustomed in quantum mechanics to dealing with Hermitian systems that properties such as the uniform convergence of the eigenspectrum $\sigma(H)$ with system size or the fact that a change in boundary conditions only leads to corrections to the spectrum of order $\mathcal{O}(1/N)$, where $N$ is the linear system size, are often accepted without questioning.

What is implicitly used is the fact that for a Hermitian matrix $H$, Weyl's inequality guarantees the stability of the eigenspectrum. Since the eigenvalues of a Hermitian matrix are real, we can order them, $\lambda_1\geq\cdots\geq\lambda_n$, where $n$ is the dimension of the matrix. For a perturbing matrix $\delta H$ we have, in particular, that
\begin{equation}
    \label{Weyl}
    |\lambda_k(H+\delta H) -\lambda_k(H)|\leq \|\delta H\|_2  
\end{equation}
where $\|.\|_2$ is the operator norm. I.e., the eigenspectrum is Lipschitz continuous on the space of Hermitian matrices with operator norm. Stability also holds in the more general case of a normal matrix, $H^\dagger H=H H^\dagger$, because a normal matrix is always diagonalizable. The normal case does include the Hermitian case but while for the Hermitian case Weyl's inequality \eqref{Weyl} provides a very stringent bound, in general only the less stringent Bauer-Fike theorem holds which, nevertheless, still guarantees the stability of the eigenspectrum. More specifically, the Bauer-Fike theorem states that for any diagonalizable Hamiltonian, $H=V\Lambda V^{-1}$ with $\Lambda$ being the diagonal matrix of eigenvalues, the following relation holds \cite{TrefethenEmbree2005}
\begin{equation}
    \label{Bauer-Fike}
    \exists \lambda \in\sigma(H): |\lambda - \mu| \leq \| V \|_2 \| V^{-1} \|_2 \|\delta H\|_2, \; \forall \mu\in\sigma(H+\delta H). 
\end{equation}
For a normal matrix $H$, the eigenvector matrix $V$ is unitary; thus $\|V\|_2=\|V^{-1}\|_2=1$ and the stability of the spectrum is guaranteed. On the other hand, for a non-normal $H$---which is what we are typically dealing with in physically relevant non-Hermitian systems---the factor $\| V \|_2 \| V^{-1} \|_2$ can become arbitrarily large because the eigenvectors no longer form an orthogonal system and can even coalesce at an exceptional point. We see that the important point is that physically interesting Hamiltonians, which are not Hermitian, are typically also non-normal. It is the fact that they are non-normal which makes the spectrum unstable, not the non-Hermiticity per se.

A useful way to visualize the spectral instability of general non-Hermitian systems is via the pseudospectrum defined as \cite{TrefethenEmbree2005,Davies2007}
\begin{equation}
    \label{pseudo}
    \sigma_\varepsilon(H) = \{ \lambda : \| (H-\lambda\mathbbm{1})^{-1} \|_2 > 1/\varepsilon \} \, .
    \end{equation}
For normal matrices the pseudospectrum is given by $\sigma_\varepsilon(H) =\{\lambda: \mbox{dist}(\lambda,\sigma(H))\leq \varepsilon \}$, i.e., the eigenspectrum $\sigma(H)$ is merely 'thickened' by all values within a range $\varepsilon$ of an eigenvalue. For non-normal matrices $H$, in contrast, the pseudospectrum contains points which can be arbitrarily far apart from any true eigenvalue. Since physically relevant non-Hermitian Hamiltonians are typically not normal, the eigenspectrum is unstable and will thus show no eigenvalue-based topological protection. It is therefore the wrong quantity to consider for topology. 

\subsection{Stability of the singular value spectrum}
What is stable instead, and can be related to the topological properties of a non-Hermitian matrix, is the singular value spectrum which is defined as
\begin{equation}
    \label{sing}
    s(H) = \{\sigma_i(H)\}_{i=1}^n,\; \mbox{with}\;\;
\sigma_i(H)=\sqrt{\lambda_i(H^\dagger H)} .
%  s(H)=\left\{\sigma: \sigma_i=\sqrt{\lambda_i(H^\dagger H)}  \right\} \, .
\end{equation}
I.e., the singular values are the square roots of the eigenvalues of the {\it Hermitian positive semidefinite matrix} $H^\dagger H$ and are therefore always stable. If $H$ itself is normal---which includes the Hermitian case---then $\sigma_i=|\lambda_i|$. This means that for Hermitian systems one can talk about the topological protection of the singular value spectrum or the eigenvalue spectrum synonymously. This has sometimes led to the misconception that topology is always related to properties of the eigenspectrum. However, this is incorrect and in general fails in the non-Hermitian case. 

We are interested in the stability of the singular value spectrum of the Hamiltonian $H$ when adding a perturbation $\delta H$. Applying Weyl's inequality \eqref{Weyl} to the Hermitian matrix $H^\dagger H$ immediately implies the inequality 
\begin{equation}
    \label{Weyl_sing}
    |\sigma_k(H+\delta H) -\sigma_k(H)|\leq \|\delta H\|_2=\sigma_{\max}(\delta H)
\end{equation}
for the singular values where $\|.\|_2$ is again the operator norm. One type of perturbation that we will consider in the following to show that topologically protected singular values are stable, is a random complex matrix $\delta H\in\mathbb{C}^{N_x\times N_y}$ with entries $\delta H_{ij}\in\{\varepsilon(x+iy):\; x,y\in[-1,1]\}$ with $\varepsilon$ characterizing the strength of the random perturbation. For such a matrix the largest singular value shows a typical scaling $\sigma_{\max}(\delta H)\sim\mathcal{O}(\varepsilon\sqrt{n})$ where $n=N_x N_y$ is the dimension of the matrix. To have a perturbation which is of constant strength when increasing the size of the matrix, the matrix elements should therefore be of order 
\begin{equation}
    \label{pert}
    |\delta H_{ij}|\sim\tilde\varepsilon=\varepsilon/\sqrt{n} \, .
\end{equation}
In the context of topological protection, we will see that what connects topological indices of the bulk with the singular value spectrum of finite truncations are K-splitting theorems. Such theorems guarantee that a certain number of singular values are separated from the bulk of the spectrum by a gap $\Delta$. The important conclusion we can already draw from the fundamental theorems above is that if the singular value spectrum has a gap $\Delta$, then this gap is stable against any perturbation $\delta H$ with
\begin{equation}
    \label{stability}
    \|\delta H\|_2=\sigma_{\max}(\delta H)<\Delta \, .
    \end{equation}
Note, in particular, that small singular values correspond to vectors that are mapped close to zero by $H$, even if no exact eigenstate at zero exists.

\subsection{Toeplitz operators and index theorems}
We are interested in parent Hamiltonians that are translationally invariant up to possible boundary terms. We note again that topology is unrelated to translational invariance and that topological features of these parent Hamiltonians will remain robust under perturbations which break it. Translational invariance should be thought of merely as a tool that allows one to calculate indices which then determine topological properties. Our goal is to use Toeplitz operator theory to fill the gap between a bulk topological classification of non-Hermitian systems and a lower bound on the protected boundary modes, thus establishing a proper bulk-boundary correspondence. 

We start with the one-dimensional case where the definitions are simplest and where Toeplitz theory was used recently to establish a bulk-boundary correspondence \cite{MonkmanSirker_nH}. We then discuss how one can generalize these definitions to the two-dimensional case.

\subsubsection{Toeplitz operators in one dimension}
For a one-dimensional translationally invariant Gaussian Hamiltonian of the form $H=\vec{c}^\dagger \mathcal{H} \vec{c}$ where $\vec{c} = (c_1\, c_2\, \cdots\, c_N)^T$ is a vector of annihilation operators on the $N$ sites of the lattice, the entries of the Hamiltonian matrix $\mathcal{H}$ only depend on the distance between the two points the creation and annihilation operators are acting on, $\mathcal{H}_{ij}=h_{i-j}$. If the Hamiltonian is restricted to the semi-infinite line, it therefore has Toeplitz form 
\begin{equation}
\label{Toplitz}
    \mathcal{H}= \begin{pmatrix}
    h_0 & h_1 & h_2 & \cdots \\
    h_{-1} & h_0 & h_1 & \cdots \\
    h_{-2} & h_{-1} & h_0 & \cdots \\
    \vdots & \vdots & \vdots & \ddots
    \end{pmatrix}
\end{equation}
where the $h_j$ are, in general, block matrices whose dimension equals the size of the unit cell. We define the symbol $h(k)$---the Fourier transform of these matrices which is also called the Bloch Hamiltonian---as
\begin{equation}
    \label{1dsymbol}
    h_j=\frac{1}{2\pi}\int_0^{2\pi}dk\, h(k)\e^{ikj} \, .
\end{equation}
Throughout this work, we assume finite-range (or sufficiently fast decaying) hopping amplitudes so that the symbol $h(k)$ is continuous in the Brillouin zone, ensuring that the associated Toeplitz operator is well defined. On the semi-infinite line, the index of a Toeplitz operator is defined as
\begin{equation}
    \label{index}
    \mbox{ind}(\mathcal{H}) = \mbox{dim}(\mbox{ker}(\mathcal{H})) - \mbox{dim}(\mbox{ker}(\mathcal{H}^\dagger)) \, .
\end{equation}
Here $\mbox{ker}(\mathcal{H})=\{v|\mathcal{H}v=0\}$ is the kernel of $\mathcal{H}$ and $\mbox{dim}$ denotes its dimension. It is important to stress that for any finite matrix $\mathcal{H}$ we have $\mbox{ind}(\mathcal{H})= 0$ because of the rank-nullity theorem (row rank is equal to column rank). A non-zero index requires an infinite-dimensional space. Gohberg's index theorem directly relates the index---the difference between right and left zero eigenvectors of a semi-infinite system---to the winding of the symbol via $\mathcal{I}=-\mbox{ind}(\mathcal{H})$ where the winding is defined as
\begin{equation}
    \label{winding}
    \mathcal{I}=\frac{1}{2\pi i}\int_0^{2\pi}dk\, \partial_k \ln\det h(k)
\end{equation}
if $\det h(k)\neq 0$. Under the assumption that $\det h(k)\neq 0$, the Toeplitz operator $\mathcal{H}$ is Fredholm. In the physics literature, this condition is called having a point gap at energy $E=0$. There are two important remarks to make here: (i) For a Hermitian system, all eigenvalues are real and there is thus no winding, $\mathcal{I}\equiv 0$. In a Hermitian system, additional symmetries are required to have non-zero topological invariants which leads to the concept of symmetry-protected topological order and to the tenfold classification. (ii) In a non-Hermitian system, there can be two types of gaps: point gaps and line gaps. The former means that for the reference energy $E$ we have $\det(h(k)-E)\neq 0$, while in the latter case there is a line of energies $E_n$ in the complex plane for which $\det(h(k)-E_n)\neq 0$ holds. It is very important to stress that having these two types of gaps is not mutually exclusive. This has important consequences for the classification of non-Hermitian topological systems. While for Hermitian systems there is only one type of gap making the tenfold classification with respect to the symmetry of such a system unique, the 38 classes found for non-Hermitian systems have to be understood as being a classification with respect to {\it both} symmetry and gap. I.e., a non-Hermitian system might belong to one class with respect to a point gap and to a different class with respect to a line gap. Accordingly, a given non-Hermitian Hamiltonian can carry distinct topological invariants associated with different choices of reference energy or spectral gap.

There is also an important difference between the case where $h(k)$ is scalar and the case where it is a matrix-valued symbol. For the scalar case, Coburn's lemma \cite{BoettcherSilbermann} states that $\mbox{ker}(\mathcal{H})$ and $\mbox{ker}(\mathcal{H}^\dagger)$ cannot both be non-zero. The sign of $\mathcal{I}$ then tells us immediately whether $\mathcal{H}$ has right or left zero modes and $|\mathcal{I}|$ equals the exact number of such modes. If $h(k)$ is a matrix symbol, on the other hand, $\mathcal{I}$ only tells us whether there are more right or more left zero modes but does not fix their total numbers. This is the same as in Hermitian Chern insulators where the Chern number also only fixes the difference between left and right propagating modes. 

So far, we have discussed the semi-infinite case. The next important step is to connect the winding number $\mathcal{I}$ with properties of a finite truncation of the Hamiltonian matrix to $L$ sites, $\mathcal{H}\to\mathcal{H}_L$. First, there is typically no uniform convergence. For Hermitian systems, we are used to the fact that if there is a vector $v$ in the kernel of $\mathcal{H}$ then there is a vector $v_L$ for the finite truncation with $\mathcal{H}_L v_L=\lambda_L v_L$ and $\lambda_L\sim \e^{-L}$. This does not hold in general in the non-Hermitian case where we instead only have
\begin{equation}
    \label{hidden}
   \lim_{L\to\infty} \| \mathcal{H}_L v_L \| =0
\end{equation}
for normalized vectors $v_L$ but $v_L$ is in general not an eigenvector. In Ref.~\cite{MonkmanSirker_nH} we have dubbed these modes 'hidden zero modes' because they cannot be found by considering the eigenspectrum of a finite Hamiltonian but instead get mapped exponentially close to zero. They only become exact eigenmodes in the semi-infinite chain. Physically, these modes are extremely long-lived metastable states. To identify these states, one has to study the singular value spectrum. This spectrum has what is called the K-splitting property if
\begin{equation}
\label{splitting}
    \lim_{L\to\infty} \sigma_n(\mathcal{H}_L)=\left\{\begin{array}{cc} 
    0 & 1\leq n\leq K \\
    >0 & n>K
    \end{array}\right. \, .
\end{equation}
For a scalar symbol $h(k)$ we have $K=|\mathcal{I}|$ while for a matrix-valued symbol, the winding number only provides a lower bound, $K\geq |\mathcal{I}|$. In higher dimensions, Toeplitz operators admit multiple inequivalent truncations corresponding to different boundary geometries. In the following, we will see that this leads to qualitatively new phenomena not present in one dimension.

\subsubsection{From one to two dimensions}
In the two-dimensional case, we can arrange the lattice sites in a one-dimensional vector while respecting the unit cell structure. Here we will assume a unit cell which stretches along the x-direction and has length 1 in the y-direction but this construction can easily be generalized. Then it is convenient to arrange the lattice sites as
\begin{equation}
    \label{lattice}
    (1,1),(2,1),\cdots, (N_x,1),\cdots,(1,N_y),\cdots,(N_x,N_y) \, .
\end{equation}
This ordering corresponds to viewing the two-dimensional lattice as a one-dimensional chain of length $N_xN_y$ with a block structure inherited from translations along the y-direction. In doing so, the Hamiltonian matrix has again the Toeplitz structure \eqref{Toplitz} with $N_y\times N_y$ entries $h_j$ where each $h_j$ is an $N_x\times N_x$ matrix which again has Toeplitz structure with elements $h_j(l)$. This structure is called block-Toeplitz with Toeplitz blocks (BTTB) or also a two-level (multi-level in general) Toeplitz matrix. We can use
\begin{equation}
    \label{symbol1}
   h_j = \frac{1}{2\pi}\int_0^{2\pi} h(k_y) \text{e}^{ik_y j}\, dk_y \, , \qquad 
   h_j(l) = \frac{1}{2\pi}\int_0^{2\pi} h_j(k_x)\text{e}^{ik_x l}\, dk_x
\end{equation}
which, when put together, leads to the two-dimensional symbol
\begin{equation}
    \label{symbol3}
    F(k_x,k_y)=\sum_{j,l} h_j(l) \text{e}^{-i(k_x l + k_y j)} \, .
\end{equation}
As in the one-dimensional case, this symbol has the dimensions of the unit cell of the system, i.e., it can be a scalar or a matrix. If there is at least a point gap for {\it all} $k_x,k_y$, which means that $\det F(k_x,k_y)\neq 0$, then the system can have a non-trivial topology. We define the size of the point gap as $\Delta_0=\min_{k_x,k_y}(|\det F(k_x,k_y)|)$.

In the scalar case, the symbol describes a mapping
\begin{equation}
\label{map}
F: T^2\longrightarrow \mathbb{C}\setminus\{0\}=S^1\times \mathbb{R}^+ 
\end{equation}
where $T^2$ is the two-dimensional torus. The homotopy class of the mapping $T^2=S^1_{k_x}\times S^1_{k_y} \longrightarrow S^1$ is characterized by two windings and thus $\sim \mathbb{Z}\oplus \mathbb{Z}$. Because we assume a point gap, the windings cannot change when varying the second parameter and we can define slice windings as
\begin{equation}
    \label{windings}
    I_x(k_y^0)=\frac{1}{2\pi i}\int_0^{2\pi} dk_x\; \partial_{k_x}\ln\det F(k_x,k_y^0) \, , \qquad
    I_y(k_x^0)=\frac{1}{2\pi i}\int_0^{2\pi} dk_y\; \partial_{k_y}\ln\det F(k_x^0,k_y) \, .
\end{equation}
Again, as long as the point gap always remains open, these slice windings are well-defined and independent of $k_x^0$ and $k_y^0$. These windings correspond to the fundamental group $\pi_1(\mathbb{C}^*)=\mathbb{Z}$ applied independently along the two cycles of the torus. They are one-dimensional in nature and do not correspond to a genuine two-dimensional bulk index. In the scalar case, the two windings \eqref{windings} fully capture the topological properties of the system. We will see later on that again a K-splitting theorem exists and that the winding numbers predict the number of edge states along the x-direction and y-direction, respectively.

If $F(k_x,k_y)$ is a matrix-valued symbol of dimension $N$, on the other hand, then we can perform a polar decomposition $F=UP$ (see also the next subsection) with $U$ unitary and $P$ a positive semidefinite Hermitian matrix. The windings \eqref{windings} then separate into a winding of the unitary matrix $U$ and a winding of $P$. Since $P$ is positive-semi-definite, its winding is trivial and we can equivalently write the windings \eqref{windings} by replacing $F(k_x,k_y)$ with $U(k_x,k_y)$. I.e., the symbol $F(k_x,k_y)$ in the general matrix-valued case describes a map $T^2\to U(N)$ and as long as we have a point gap, $\det F(k_x,k_y)\neq 0$, the bulk invariants are still the two windings $I_x$ and $I_y$ as given in Eq.~\eqref{windings}. As long as there is no symmetry-protected decomposition of $F(k_x,k_y)$ into invariant subspaces, there is no intrinsic two-dimensional bulk invariant, $\pi_2[U(N)]=0$. However, if the system has a line gap then one can define a projection onto one of the bands and this projected unitary $U_P(N)$ defines a vector bundle, which can be characterized by a Chern number
\begin{equation}
    \label{Chern}
    \nu =\frac{1}{2\pi i}\int_0^{2\pi} dk_x \int_0^{2\pi} dk_y \Tr(U_P^{-1}\partial_{k_x}U_P \; U_P^{-1}\partial_{k_y}U_P) \, .
\end{equation}
Up to a rotation that does not change the topology of the system, we can choose the line gap to be along the real axis. We can then, again without changing the topology, flatten the bands to make them purely real. I.e., the topological characterization with respect to a line gap is identical for Hermitian and non-Hermitian systems. In other words, the non-Hermitian two-dimensional Chern insulator is just a topologically trivial deformation of a Hermitian one. We will thus not consider this trivial extension any further in this article.

In Sec.~\ref{Sec_gen}, we will connect these fundamental definitions of the bulk invariants in the two-dimensional case with the properties at boundaries. In two dimensions, different choices of truncation correspond to different boundary geometries and we will analyze how these distinct geometries enter the bulk–boundary correspondence.

\subsection{Decompositions}
When discussing matrix-valued symbols, we have seen that a polar decomposition is useful and that the symbol ultimately describes a mapping into the space of unitary matrices. Here we want to recall the main definitions of polar and singular value decompositions and show that the unitary matrices obtained from a matrix-valued symbol are stable objects for a topological classification.

If $F$ is a non-Hermitian matrix, then we can use the singular value decomposition
\begin{equation}
    \label{SVD}
    F=U_s\Sigma V_s^\dagger
\end{equation}
where $\Sigma=\mbox{diag}(\sigma_1,\cdots,\sigma_N)$ is the diagonal matrix of singular values and $U_s$ and $V_s$ are unitary. In a polar decomposition, we have instead
\begin{equation}
    \label{polar}
    F=UP
\end{equation}
with $U$ unitary and $P$ positive semidefinite and Hermitian. 

These two fundamental definitions imply a number of useful relations. First of all, $F^\dagger F=P^\dagger U^\dagger U P=P^\dagger P=P^2$
which implies that 
$P=(F^\dagger F)^{1/2}=(V_s\Sigma U_s^\dagger U_s\Sigma V_s^\dagger)^{1/2}=(V_s\Sigma^2V_s^\dagger)^{1/2}=V_s\Sigma V_s^\dagger $. For the unitary $U$ in the polar decomposition it follows that
\begin{equation}
    \label{unitary_relation}
    U=F P^{-1}=U_s\Sigma V_s^\dagger V_s\Sigma^{-1}V_s^\dagger=U_sV_s^\dagger
\end{equation}
which is just the product of the two unitary matrices appearing in the singular value decomposition. Importantly, if $F$ is invertible and has a point gap, i.e., if its smallest singular value is bounded from below by a strictly positive constant, then the polar decomposition is unique and the unitary matrix $U$ depends continuously on $F$. In this case, small perturbations of $F$ lead to small perturbations of $U$. The polar unitary $U(k_x,k_y)$ is therefore a stable object that can be used to define bulk topological invariants. This stability is lost if the point gap closes, reflecting the fact that the polar decomposition becomes ill-defined when singular values approach zero.

\section{General results}
\label{Sec_gen}
In this section we present our general results for the bulk-boundary correspondence in topological two-dimensional non-Hermitian systems based on Toeplitz operator theory. Going from a bulk to a boundary problem, there are two main ways in which the system can be truncated. One possibility is that we only have a single boundary which stretches all the way either along the x-direction or the y-direction. This is the case where the Toeplitz operator has been compressed to the infinite half plane. For a finite system, we have to use periodic boundary conditions along one direction and open boundary conditions along the other, creating a finite-size proxy of the half plane with a tube geometry. This is very similar to the one-dimensional case \cite{MonkmanSirker_nH} where one studies a finite chain with open ends at both sides and connects these finite-size results with the case of the semi-infinite chain. The other main possibility is to have a semi-infinite quarter plane with two edges which are joined at a corner. In this scenario we might have stable corner modes in addition to edge modes. One can further separate this case into two subclasses: In the first, the edges are gapless and edge modes and potential corner modes might co-exist. In the second, not only the bulk but also the edges are gapped, and the bulk-boundary correspondence is then between bulk invariants and a finite number of corner modes \cite{Simonenko1968,DouglasHowe1971}. For a finite system, one considers in both cases a finite square with open boundaries along all its edges. Finally, we should also mention that we are here only considering the quarter plane which has a rectangular-shaped corner. In principle, more general corners which can either be concave or convex are also possible \cite{Hayashi2019Toeplitz}. 

\subsection{The half plane case}
We start with the simpler half-plane case, i.e., the case where the Toeplitz operator only has a single straight boundary. The goals are (i) to use the bulk invariant to set a lower bound for the number of protected edge modes along this boundary, and (ii) to use K-splitting theorems to predict how these protected edge modes can be found by investigating the finite-size singular value spectrum.

W.l.o.g.~we take periodic boundary conditions along the x-direction and open boundary conditions along the y-direction. The Hamiltonian in real space is then a $N_y\times N_y$ Toeplitz matrix \eqref{Toplitz} with $N_x\times N_x$ blocks $h_j$. I.e., our matrix is of BTTB form and we denote its corresponding symbol by $F(k_x,k_y)$, see Eq.~\eqref{symbol3}. As required for non-trivial topology, we assume $\det F(k_x,k_y)\neq 0$. From the block symbol $h(k_y)$, see Eq.~\eqref{symbol1}, we can then define a winding number $\mathcal{I}$ as in the one-dimensional case \eqref{winding}. To calculate the winding number $\mathcal{I}$ we can use that $k_x=2\pi n/N_x$ is a good quantum number so that
\begin{equation}
    \label{det}
   \det h(k_y)=\prod_{n=0}^{N_x-1} \det F\left(\frac{2\pi}{N_x
   }n,k_y\right) \, .
\end{equation}
This allows us to write the winding number as
\begin{equation}
\label{HN_winding}
    \mathcal{I} = \frac{1}{2\pi \im}\int_0^{2\pi}dk_y \, \partial_{k_y} \ln\det h(k_y)  
    =
\frac{1}{2\pi \im}\sum_{n=0}^{N_x-1}\int_0^{2\pi} dk_y\, \partial_{k_y} \ln \det F\left(\frac{2\pi}{N_x}n,k_y\right) 
= N_x I_y 
\end{equation}
where the slice winding number $I_y$ is defined in Eq.~\eqref{windings} and $I_y$ cannot depend on $k_x$ because the point gap never closes, $\det F(k_x,k_y)\neq 0$. The sign of the slice winding determines on which boundary the edge state is localized.

For the semi-infinite case where we only have one boundary either at the top or the bottom of the tube and send the other boundary to infinity, the index theorem \eqref{index} is applicable and we find
\begin{equation}
    \label{index2}
    N_xI_y = \mbox{dim}(\mbox{ker}(\mathcal{H}^\dagger)) - \mbox{dim}(\mbox{ker}(\mathcal{H})) \, .
\end{equation}
This means that in a topologically non-trivial phase with $I_y\neq 0$ we have an entire family of edge eigenmodes with zero energy in the semi-infinite system whose number scales linearly with $N_x$. Here it is again important to distinguish between the cases where the symbol $F(k_x,k_y)$ is scalar and the case where it is matrix valued. In the scalar case, Coburn's lemma is applicable and $\mbox{ker}(\mathcal{H})$ and $\mbox{ker}(\mathcal{H}^\dagger)$ cannot be non-zero at the same time. Thus $|I_y|$ predicts the exact number of zero modes and the sign predicts at which boundary they live. In the matrix-valued case, on the other hand, both kernels can be non-zero at the same time and the bulk invariant only predicts the difference in the number of zero modes when comparing the two semi-infinite tubes---one with a boundary at the top and the other with a boundary at the bottom---but no longer their total numbers. Note that this is not a flaw of the theory presented here but rather a fundamental property of index theorems. The perhaps best known case is the Hermitian Chern insulator where the Chern number also only predicts the difference between right and left propagating edge modes, which determines the conductivity, but not their total numbers.

For a finite system with boundaries at the top and the bottom of the tube, there will typically be no edge eigenmodes with exponentially small eigenvalues---despite the existence of exact kernel modes in the semi-infinite limit---because there is in general no uniform convergence of the eigenvalue spectrum with system size in a non-Hermitian system. The edge modes are usually hidden in a finite system and correspond to vectors which get mapped by the Hamiltonian exponentially close to zero, see Eq.~\eqref{hidden}, but which are not part of the eigenspectrum. Instead, the bulk slice windings and the exact zero modes in the semi-infinite system are related to exponentially small singular values in the finite system, separated from the bulk spectrum by a gap. Because we have two boundaries in a finite system, we have to consider, in addition to $\mathcal{H}$, also the reflected Hamiltonian $\tilde {\mathcal{H}}$ obtained from $\mathcal{H}$ by the replacement $h_j\to h_{-j}$. We define $K$ by the kernels of two semi-infinite tubes; one with its boundary at the top, the other with the boundary at the bottom \cite{BoettcherSilbermann}
\begin{equation}
    \label{K}
    K = \mbox{dim}(\mbox{ker}(\mathcal{H})) + \mbox{dim}(\mbox{ker}(\tilde{\mathcal{H}}))
\end{equation}
and the singular value spectrum then has the K-splitting property \eqref{splitting} with $K$ as defined above. Crucially, the K-splitting is stable and holds also for the finite tube. Now we have to again distinguish the cases where the symbol is scalar and the one where it is matrix valued. In the scalar case, $\tilde{\mathcal{H}}=\mathcal{H}^T$ and $\mbox{dim}(\mbox{ker}(\mathcal{H}^T))=\mbox{dim}(\mbox{ker}(\mathcal{H}^\dagger))$ so that we can write 
\begin{equation}
    \label{K2}
    K=N_x |I_y|=|\mbox{dim}(\mbox{ker}(\mathcal{H}^\dagger)) - \mbox{dim}(\mbox{ker}(\mathcal{H}))| \, .
\end{equation}
So in the scalar case there are exactly $N_x |I_y|$ singular values which are separated from the bulk spectrum by a gap and which will go to zero when increasing the length of the tube. The corresponding singular vectors are exponentially localized at either end of the tube. If $F(k_x,k_y)$ is matrix valued, then $\tilde{\mathcal{H}}$ is different from the transpose and we only know that there will be at least $K\geq N_x |I_y|$ singular values which will go to zero \cite{MonkmanSirker_nH}. 

\subsection{The quarter plane case}
In the quarter plane case, we want to consider a semi-infinite plane with two boundaries that meet at a right angle and its finite truncations to rectangles with open boundaries along all four edges. As in the half plane case, we can have families of edge modes, however the physics can become more complicated for two reasons: (1) If there are families of edge modes along edges which are joined at a corner then they can hybridize which means that even in the scalar case, their total number is not necessarily the number of modes along one edge plus the number of modes along the other edge. (2) In addition to the family of edge states, there can also be corner modes whose support is concentrated at the corner where the two edges meet. A very different case is realized if not only the bulk but also the edges are gapped. Like in the Hermitian case, we might then still have a higher-order topological phase---not characterized by the slice windings $I_{x,y}$ which will be trivial in this case---which induces a bulk-boundary correspondence where only $\mathcal{O}(1)$ gapless corner modes exist. 

\subsection{Gapless edges}
\label{Quarter_gapless}
If either of the slice windings $I_x$ or $I_y$ is non-zero, then gapless edge modes do exist. There are two main cases to distinguish: (i) Only one of the slice windings is non-zero, (ii) both are non-zero. In the first case, nothing fundamentally changes compared to the half-plane case. We can think of this case as starting from the tube with the non-zero slice winding along the open boundary direction. Then, slicing open the tube introduces an additional boundary but this boundary is fully gapped because the slice winding along this direction is zero. Therefore the new boundary does not interact with the existing ones and the results from the half-plane case hold. W.l.o.g. we can assume $I_y\neq 0$ and $I_x=0$. Then the quarter plane will have a family of $K=N_x |I_y|$ edge modes along the y-boundary if the symbol is scalar and $K\geq N_x|I_y|$ if it is matrix valued.

The more interesting and complicated case is if both slice windings are non-zero. Then, there are families of edge modes along both the x-boundaries and y-boundaries and because these boundaries are connected at the corners, these families of edge modes will interact with each other. Therefore the naive expectation that the total number of edge modes is bounded by $K\geq N_x |I_y| + N_y |I_x|$ is in general incorrect. The other question is whether the edge modes hybridize to form modes which live on both edges
\begin{equation}
    \label{edge}
    |\Psi_{ij}| \sim |z|^i + |w|^j
\end{equation}
with $|z|<1$ and $|w|<1$ or whether the result is corner modes
\begin{eqnarray}
    \label{corner}
    |\Psi_{ij}|\sim |z|^i |w|^j 
\end{eqnarray}
which exponentially decay in both directions at the same time. Here it is important to note that each of the two families of edge modes (the one localized in the x-direction and the one localized in the y-direction) is the solution of separate recursion relations. Furthermore, because each member of the family of edge modes decays exponentially away from the boundary it is localized at, a significant hybridization between the edge mode families will happen only in a small region near the corner, while the modes are unchanged away from the corner. Therefore a generic hybridization will lead to the additive structure \eqref{edge}. The multiplicative structure \eqref{corner}, on the other hand, will require a recurrence relation which simultaneously enforces an exponential decay along both directions which is non-generic and is only possible if the Hamiltonian has additional structure such as a factorization of the symbol or an equivalent constraint enforcing simultaneous annihilation of the two boundary transfer operators. The most important point to note is that even if corner modes do exist in this scenario, they are not protected by a Fredholm corner index theorem in the sense of Hayashi \cite{Hayashi2019Toeplitz,Hayashi2023QuarterPlane} because this would require the edges to be gapped as well. On the other hand, that does not mean that such corner modes are generically unstable. If they do exist, then the corresponding singular values typically decay faster with system size than those belonging to edge modes because the residual support of these modes along a finite-size boundary is exponentially decaying along the boundary while it is of the same order for all boundary sites for an edge mode. As a result, there will be a gap between the singular values belonging to corner modes and those belonging to edge modes which, in turn, are separated by another gap, which is topologically protected, from the bulk modes. Perturbations which respect the structure responsible for the corner modes and which are small compared to the gap to the edge modes will thus leave the corner modes stable. The stability of such corner modes is thus spectral rather than index-theoretic: it relies on a hierarchy of singular-value gaps rather than on Fredholm stability. To summarize, having both slice windings non-zero leads to edge modes localized on both edges and is also a necessary but not sufficient condition to have corner modes in a model with gapless edges. These modes are, however, only realized if additional restrictions are present in the Hamiltonian but, if present, can also be stable against perturbations. 

\subsection{Gapped edges}
Clean corner modes in an index-theoretic sense are only possible if not only the bulk but also the edges of the quarter plane operator are gapped, making the quarter plane operator Fredholm with closed range \cite{Simonenko1968,DouglasHowe1971}. This corresponds to the case where a point gap is present, $\det F(k_x,k_y)\neq 0$, and the two slice windings vanish, $I_x=I_y=0$. In this regime, no families of edge-localized zero modes exist in the half plane geometry and the kernel and co-kernel of the corner plane operator are finite, making it possible to define an index using Eq.~\eqref{index} with $\mathcal{H}$ now interpreted as the quarter-plane compression of the Toeplitz operator.

Index theorems relating this corner index to bulk data have been established in the mathematical literature, most notably by Hayashi \cite{Hayashi2019Toeplitz,Hayashi2023QuarterPlane,Hayashi2021Classification}. These results provide a bulk–corner correspondence in the strict sense: the number of corner-localized zero modes is fixed by bulk invariants and cannot be changed without closing either the bulk gap or an edge gap. In Hermitian systems, this framework underlies the theory of higher-order topological insulators. It is important to note that this index is fundamentally different from the slice windings $I_x,I_y$ because it is genuinely two-dimensional.

Our focus in this work is on showing that these known index theorems also apply in the non-Hermitian setting and that they do not require crystalline symmetries that were central in the original topological classification based on quantized polarizations \cite{Benalcazar2017Science}. Furthermore, we will concentrate on connecting such indices with the singular value spectrum and the corresponding corner modes and demonstrate their stability against disorder which respects the symmetry responsible for the topological order. The non-Hermitian BBH model discussed in Sec.~\ref{Sec_BBH} provides a concrete physical realization of this scenario. We also note that the singular value framework allows one to clearly distinguish between the case where a corner mode is protected by an index---which is the case where we have a finite number of singular values separated by a gap from the bulk spectrum---and the case where such modes coexist with families of edge modes. In the latter case, there is a gap between corner and edge modes which is not protected by an index and a further gap to the bulk which is protected by the index corresponding to the slice windings. I.e., in this case the corner mode is only spectrally protected.

\section{Two-dimensional Hatano-Nelson model}
\label{Sec_HN}
\begin{figure*}[htp]
    \centering
    \includegraphics[width=0.48\linewidth]{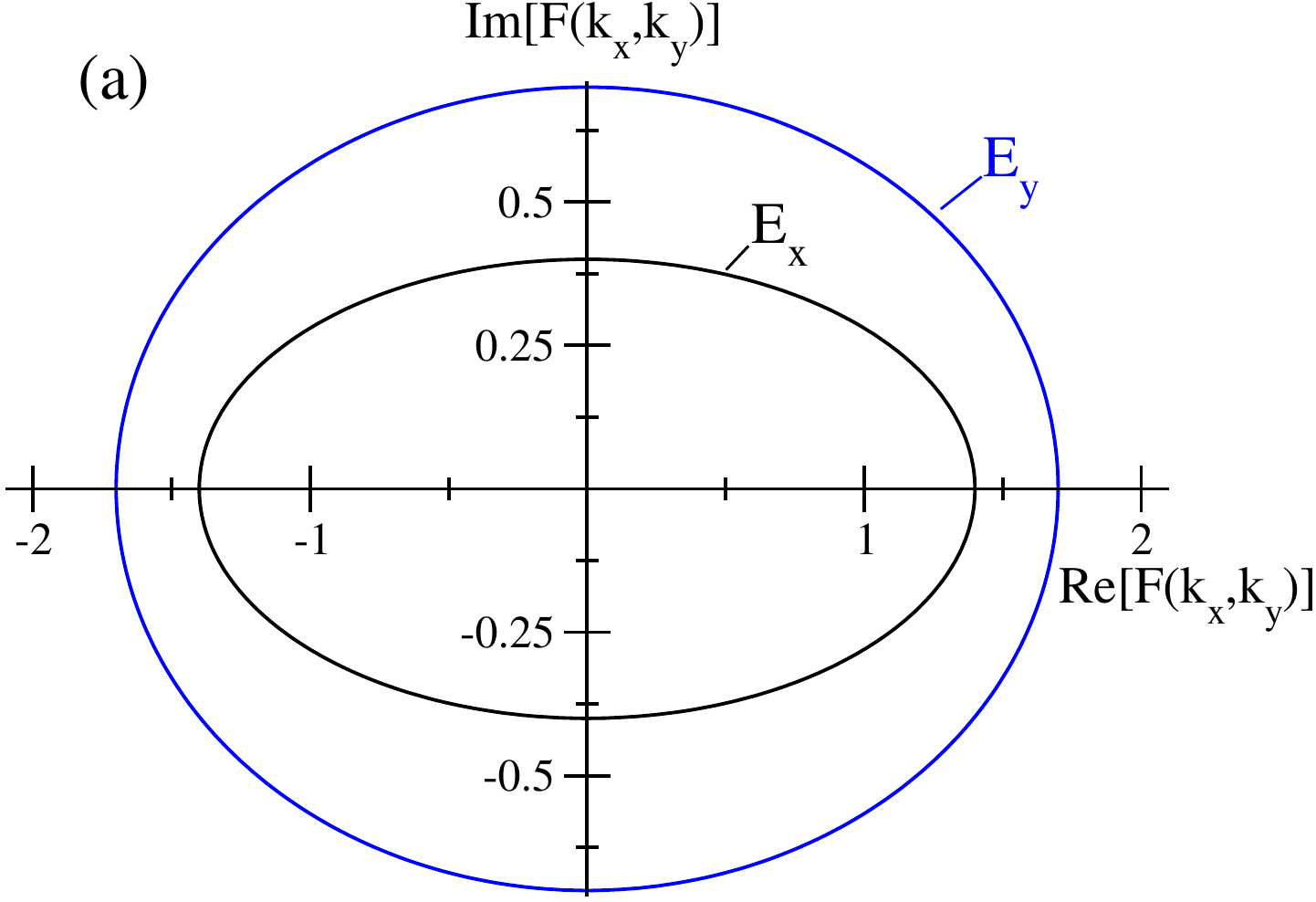}
    \includegraphics[width=0.48\linewidth]{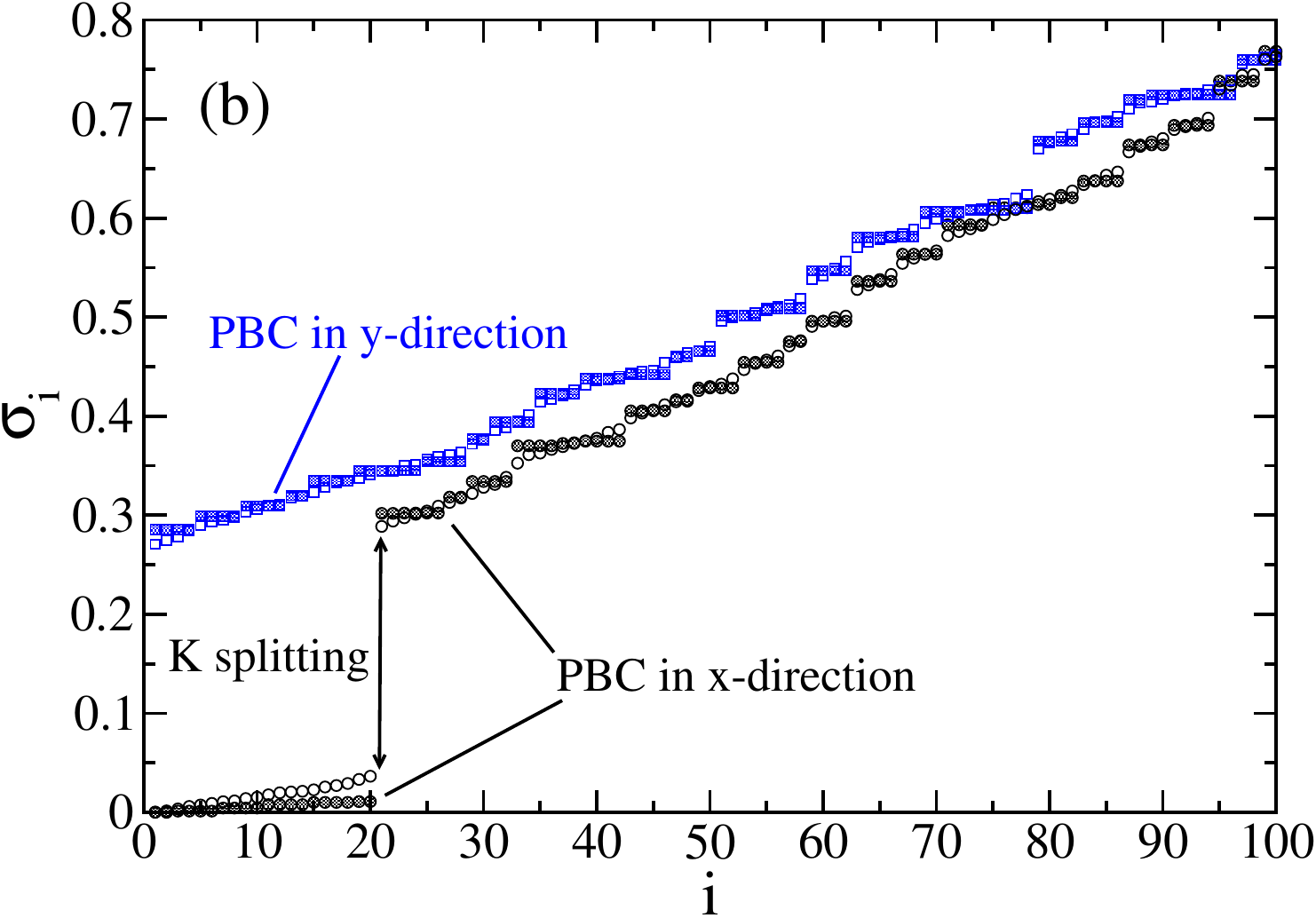}
    \includegraphics[width=0.48\linewidth]{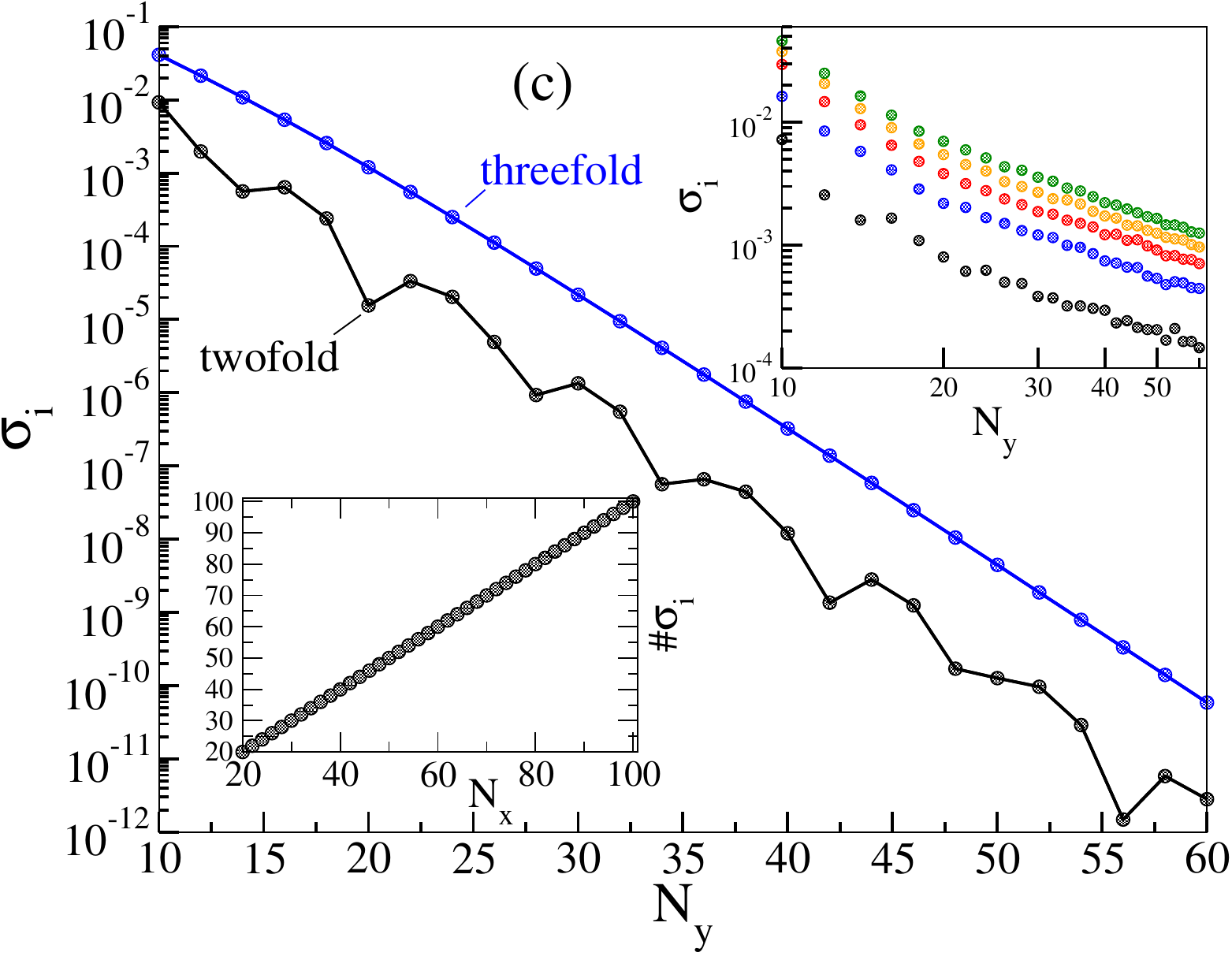}
    \includegraphics[width=0.48\linewidth]{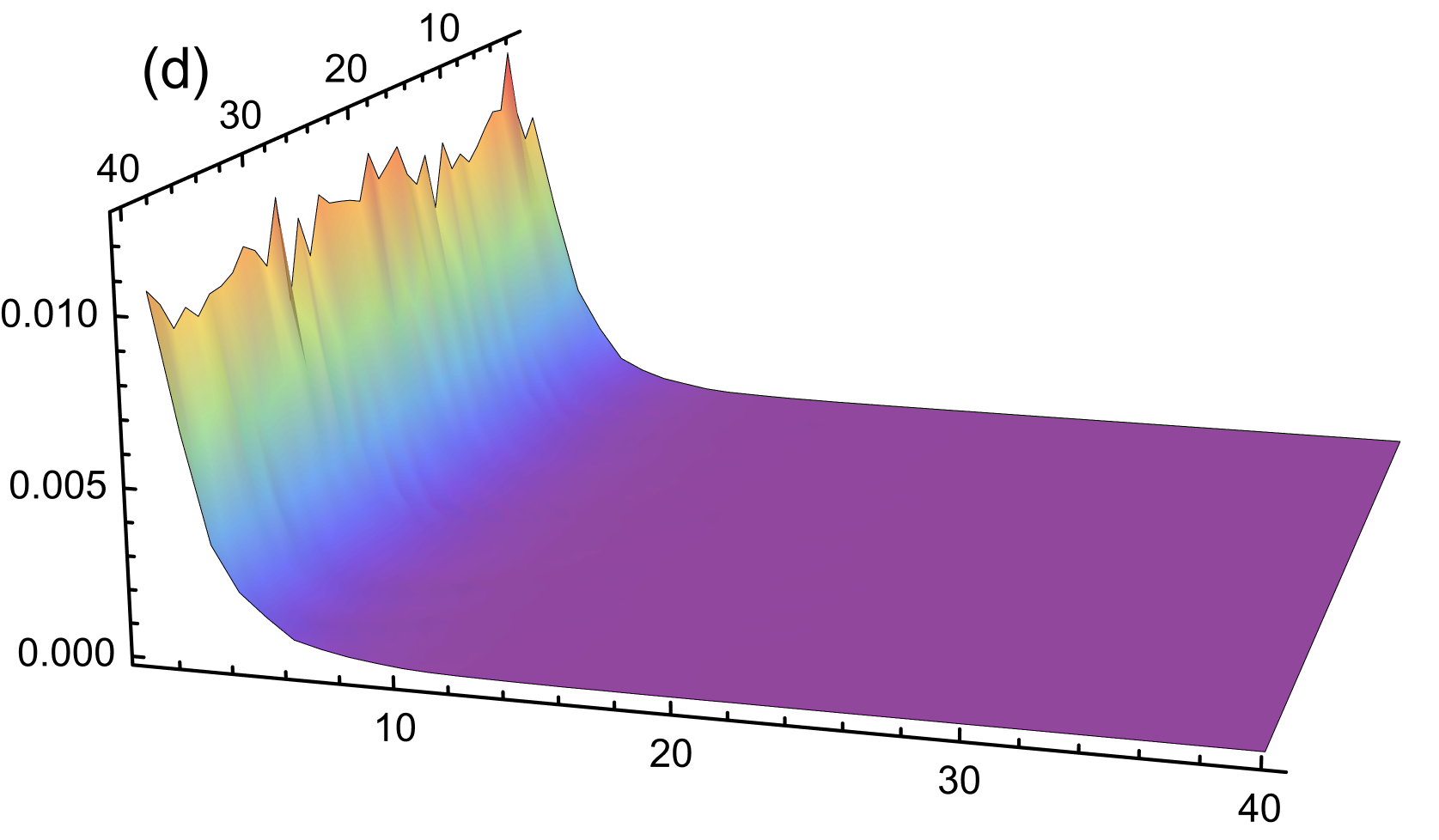}
    \caption{Hatano-Nelson model \eqref{HN} with parameters $t_R=0.5$, $t_L=0.9$, $t_U=1.2$, $t_D=0.5$ corresponding to a point gap $\Delta_0\approx 0.26$ and windings $(I_x,I_y)=(0,-1)$: (a) $F(k_x,k_y)=E_x+E_y$. (b) First $100$ singular values for a $20\times 20$ lattice. The filled symbols represent the unperturbed system, the open symbols a perturbed system with $\varepsilon=0.1$, see Eq.~\eqref{pert}. The perturbation destroys degeneracies but the spectrum is stable. (c) Main: Scaling of five smallest singular values for a clean $N_x=N_y$ system. Upper inset: Same for a perturbed system with $\varepsilon=0.1$, averaged over $20$ realizations. Lower inset: Number of topologically protected singular values as a function of $N_x$ for $N_y=20$. (d) $|\Psi(x,y)|^2$ for the singular vector belonging to the smallest singular value of a perturbed $40\times 40$ system with $\varepsilon=0.1$ averaged over $100$ samples showing stable edge localization.}
    \label{Fig_HN}
\end{figure*}

We now want to illustrate our general results using a few physically relevant examples. As an example of a model with a scalar symbol $F(k_x,k_y)$ we consider first a two-dimensional generalization of the Hatano-Nelson model \cite{HatanoNelson}
\begin{equation}
    \label{HN}
    H=\sum_{i,j} \left[t_R c_{i+1,j}^\dagger c_{i,j}+t_L c_{i,j}^\dagger c_{i+1,j}+t_Dc_{i,j+1}^\dagger c_{i,j}+t_U c_{i,j}^\dagger c_{i,j+1}\right]
\end{equation}
which lives on a $N_x\times N_y$ lattice with non-reciprocal hopping amplitudes $t_{R/L}$ and $t_{D/U}$. We can construct the corresponding Hamiltonian matrix $\mathcal{H}$ using the ordering of the sites \eqref{lattice} and by defining the $N_x\times N_x$ shift operator 
\begin{equation}
    \label{shift}
    S_{N_x}=\begin{pmatrix}
0 & 1 & 0 & 0 &\cdots \\
0 & 0 & 1 & 0 &\cdots \\
0 & 0 & 0 & 1 &\cdots \\
\vdots & \vdots & \vdots & \ddots & \\
(1) & 0 & \cdots & \cdots & 0
\end{pmatrix}
\end{equation}
and similarly for $S_{N_y}$. The "1" in the bottom left corner has to be added to realize periodic boundary conditions either in $x$ or $y$ or even in both directions. Then we can write
\begin{equation}
    \label{HN2}
    \mathcal{H}= \mathbb{I}_{N_y}\otimes (t_L S_{N_x}+t_R S^T_{N_x})+ (t_U S_{N_y} +t_D S_{N_y}^T)\otimes \mathbb{I}_{N_x}.
\end{equation}
The resulting matrix has BTTB form, see Eq.~\eqref{Toplitz}, where each $h_j$ block is an $N_x\times N_x$ matrix with $h_0=t_L S_{N_x}+t_R S^T_{N_x}$, $h_1=t_U\mathbb{I}_{N_x}$, $h_{-1}=t_D\mathbb{I}_{N_x}$, and all other blocks equal to zero. Using \eqref{symbol1} we can define the $N_x\times N_x$ block symbol
\begin{equation}
    \label{HN3}
    h(k_y)= h_0 + h_1 \e^{-ik_y} + h_{-1} \e^{ik_y}
\end{equation}
and using the second Fourier transform \eqref{symbol3} we obtain the scalar symbol
\begin{equation}
    \label{HN4}
    F(k_x,k_y)=t_L \e^{-ik_x}+t_R \e^{ik_x} +t_U \e^{-ik_y}+t_D\e^{ik_y} \, .
\end{equation}
We can now ask what winding numbers $I_x$ and $I_y$ are possible if $\det F(k_x,k_y)\neq 0$ for all $k_x$ and $k_y$, i.e., if we have a point gap at zero. We can write the symbol as
\begin{equation}
    \label{ellipses}
    F(k_x,k_y) = E_x + E_y 
    = (t_R + t_L) \cos k_x + i(t_R-t_L)\sin k_x 
    + (t_D + t_U) \cos k_y + i(t_D-t_U)\sin k_y 
\end{equation}
which describe two ellipses, see Fig.~\ref{Fig_HN}(a): $E_x$ with semi-axes $|t_R+t_L|$ and $|t_R-t_L|$ and $E_y$ with semi-axes $|t_D+t_U|$ and $|t_D-t_U|$. We are explicitly excluding here the degenerate cases $t_R=\pm t_L$ or $t_D=\pm t_U$ which have no point gap and are therefore topologically trivial. Then, there are three possibilities: 1) $E_y\subset E_x$: Then $I_x=\pm 1$ and $I_y=0$ and the gap never closes. 2) $E_x\subset E_y$: Then $I_y=\pm 1$ and $I_x=0$ and again the gap never closes. 3) $E_x$ and $E_y$ intersect. Then also $E_x$ and $-E_y$ ($\pi$ rotated ellipse) intersect which means there is a point with $F(k_x,k_y)=0$ and thus the point gap closes. With an open point gap the only possible windings are therefore $(I_x,I_y)=(\pm 1,0)$ and $(I_x,I_y)=(0,\pm 1)$.

We now turn to the main goal of our work which is to connect the bulk invariants to the number of protected boundary modes. We start by considering the tube case where we have periodic boundary conditions in x-direction and open boundary conditions in y-direction which is the finite-dimensional proxy for the half plane. Then our Hamiltonian matrix \eqref{HN2} is a $N_y\times N_y$ Toeplitz matrix with $N_x\times N_x$ blocks $h_j$. From the block symbol $h(k_y)$, see Eq.~\eqref{HN3}, we can then define a winding number $\mathcal{I}$ as in the one-dimensional case \eqref{winding}. The K-splitting theorem for block Toeplitz matrices is directly applicable and predicts that for a scalar symbol there are exactly $K = |\mathcal{I}|$ topologically protected singular values $\sigma_i$ with $\lim_{N_y\to\infty} \sigma_i =0$. The corresponding singular vectors belong to boundary modes which live either at the upper or at the lower cut of the tube depending on the sign of $\mathcal{I}$. To calculate the winding number $\mathcal{I}$ we can use Eqs.~(\ref{det}, \ref{HN_winding}) which yields $\mathcal{I}=N_x I_y$
with $I_y\in \{-1,0,1\}$. I.e., in a topologically non-trivial phase, $|I_y|\neq 0$, we find a family of protected singular values and corresponding edge modes, whose number scales linearly with $N_x$. One can therefore interpret $I_y$ as a constant index density along the edge in the sense that each transverse momentum sector contributes one protected boundary mode. Of course, one can do the same construction by interchanging x and y in the construction of the Toeplitz matrix which will then tell us about the number of edge modes in the x-direction. Note, however, that for this model at least one of the two slice windings is always zero. I.e., we can never have stable edge modes along both directions at the same time. 

In Fig.~\ref{Fig_HN} results for the two-dimensional Hatano-Nelson model in a phase with $(I_x,I_y)=(0,-1)$ are shown. The symbol $F(k_x,k_y)=E_x(k_x)+E_y(k_y)$ consists of two ellipses, see Fig.~\ref{Fig_HN}(a), and $E_x$ lies entirely inside $E_y$ consistent with the windings and also showing that a point gap is always open. The singular value spectrum of the unperturbed system shown in Fig.~\ref{Fig_HN}(b) has $N_x$ protected singular values if we choose PBC in the x-direction and OBC in the y-direction and none for PBC in the y-direction and OBC in the x-direction, fully consistent with the K-splitting theorem.
For PBC in x-direction, the $N_x$ protected singular values scale exponentially to zero with $N_y\to\infty$ in the unperturbed system. If we perturb the system by a random complex matrix, then the protected singular values after averaging over different realizations still scale to zero but the scaling is now polynomial, see Fig.~\ref{Fig_HN}(c). The corresponding singular vectors represent edge states which are completely stable against perturbations smaller than the gap $\Delta_0=\min_{k_x,k_y} |F(k_x,k_y)|$, see Fig.~\ref{Fig_HN}(d). These states become exact zero-energy eigenstates of the semi-infinite Toeplitz operator. 

\begin{figure*}
    \centering
    \includegraphics[width=0.32\linewidth]{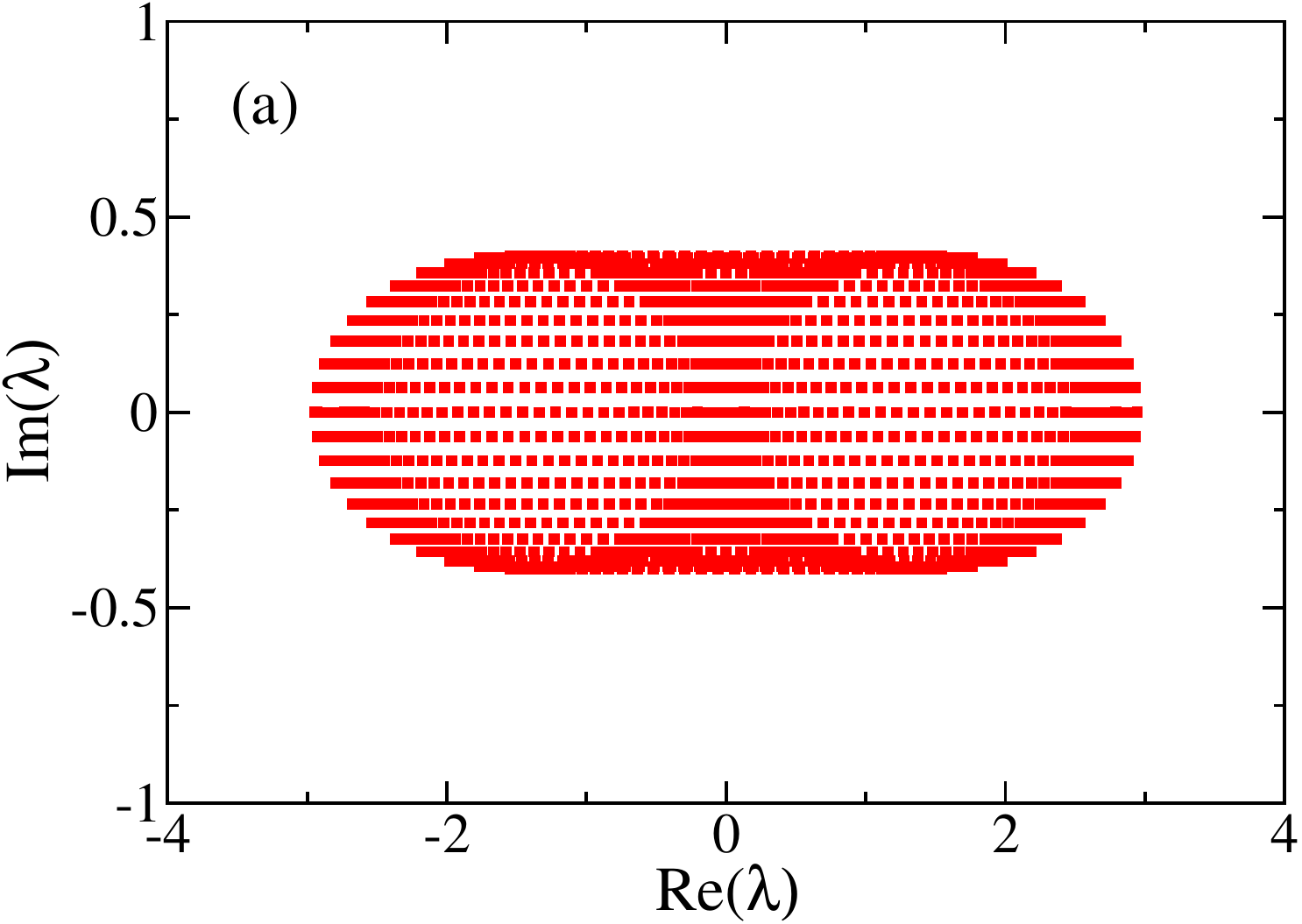}
    \includegraphics[width=0.32\linewidth]{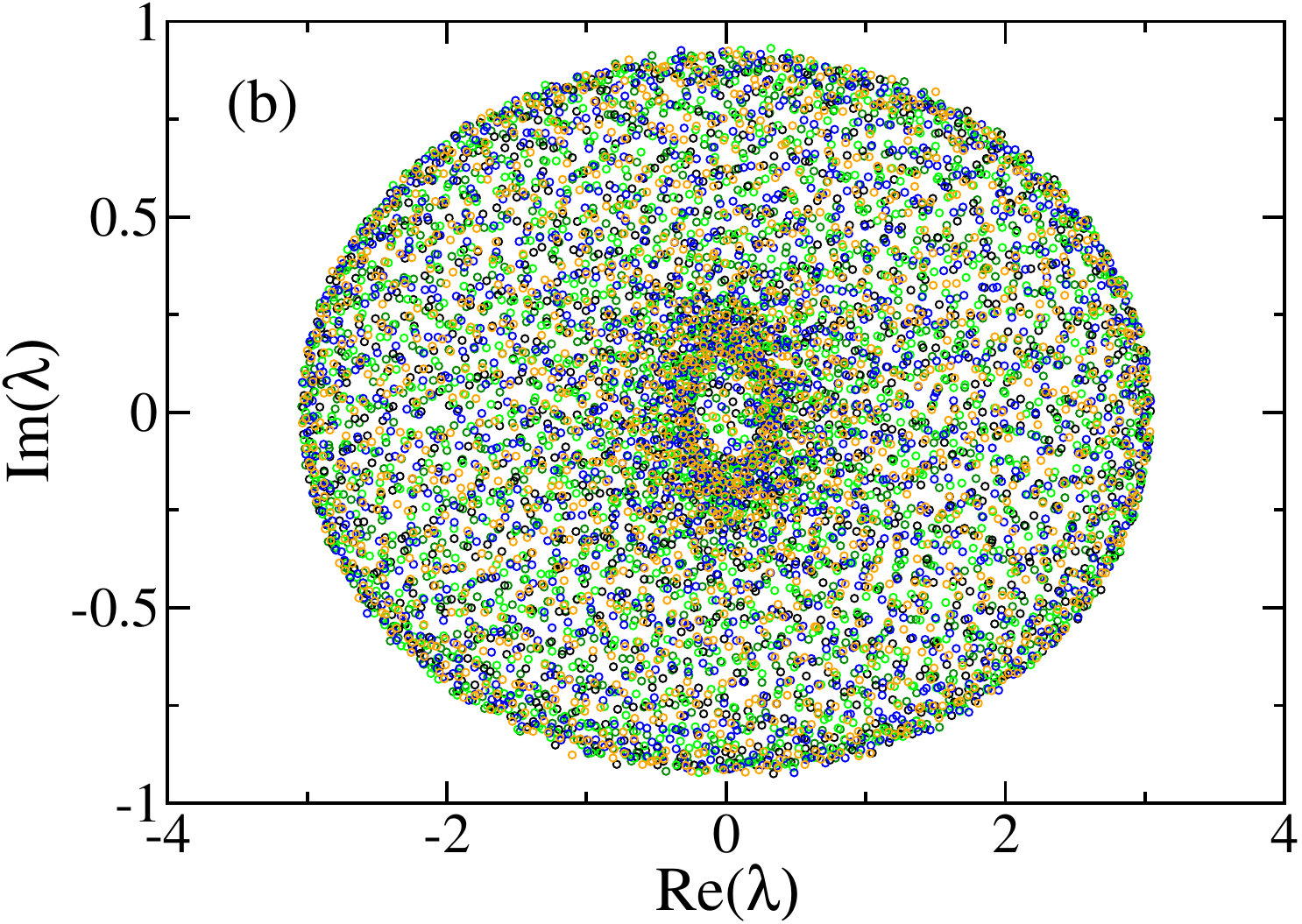}
     \includegraphics[width=0.32\linewidth]{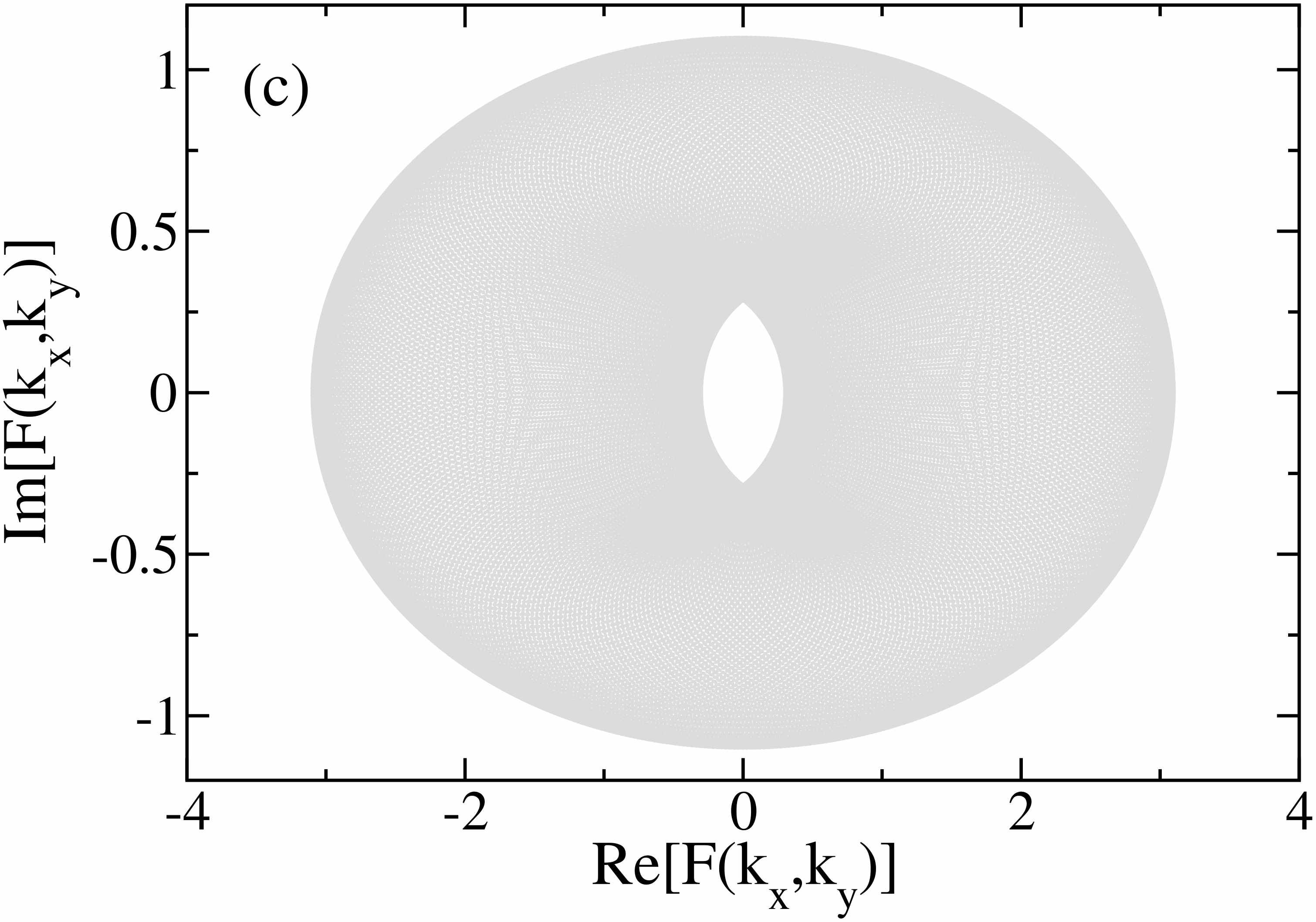}
    \caption{(a) The eigenspectrum of the unperturbed $20\times 20$ Hatano-Nelson model \eqref{HN} with parameters as in Fig.~\ref{Fig_HN}, and (b) the spectrum for 5 realizations of the perturbed system with $\varepsilon=0.1$. (c) For $N_x,N_y\to\infty$ the eigenspectrum of the perturbed system will fill out the entire image of $F(k_x,k_y)$ for any arbitrarily small perturbation $\varepsilon$ \cite{TrefethenEmbree2005}.}
    \label{Fig_HN2}
\end{figure*}

This has to be contrasted with the behavior of the eigenspectrum shown in Fig.~\ref{Fig_HN2}.
The spectrum of the unperturbed system, see Fig.~\ref{Fig_HN2}(a), is unstable and drastically modified by small perturbations, see Fig.~\ref{Fig_HN2}(b). It is obvious that none of the eigenvalues are stable. In fact, for $N_x,N_y\to\infty$ it is well known that, for any fixed perturbation strength $\varepsilon>0$, the pseudospectrum of $\mathcal{H}$, see Eq.~\eqref{pseudo}, converges to the image of the symbol $F(k_x,k_y)$ \cite{TrefethenEmbree2005}, see Fig.~\ref{Fig_HN2}(c). Since this image has a finite point gap $\Delta_0=\min_{k_x,k_y}|F(k_x,k_y)|>0$ around zero, the spectrum of the perturbed system contains no eigenvalues inside this gap in the thermodynamic limit. This explicitly demonstrates that a topological framework based on the eigenvalue spectrum fails.

\subsection{Recurrence relation}
It is instructive to directly derive the edge states for the semi-infinite tube from a recurrence relation and to show explicitly how the bulk-boundary correspondence arises. We consider again a system with PBC in the x-direction and OBC in the y-direction. Then, $k_x$ is a good quantum number and we can perform a partial Fourier transform
\begin{equation}
    \label{Fourier}
    c_{i,j}=\frac{1}{\sqrt{N_x}}\sum_{k}\e^{-\im k i} c_{k,j} \, .
\end{equation}
The Hamiltonian is then of the form $H=\sum_{k}H_{k}$ with
\begin{equation}
    \label{Hk}
    H_{k} =\sum_j \left( \mu_k c^\dagger_{k,j}c_{k,j}+t_D c^\dagger_{k,j+1}c_{k,j}+t_U c^\dagger_{k,j}c_{k,j+1}\right) \, .
\end{equation}
and $\mu_k=t_R\e^{\im k}+t_L\e^{-\im k}$. I.e., for each $k$ value we now have a one-dimensional problem to solve. We are, in particular, interested in zero-energy edge states for a system which is semi-infinite in the y-direction. This means we have to solve the matrix equation $\mathcal{H}_k\mathbf{\Psi}=0$ with the tri-diagonal matrix
\begin{equation}
    \label{Hk2}
    \mathcal{H}_k=\begin{pmatrix}
\mu_k & t_U & 0 & 0 &\cdots \\
t_D & \mu_k & t_U & 0 &\cdots \\
0 & t_D & \mu_k & t_U &\cdots \\
\vdots & \vdots & \vdots & \ddots 
\end{pmatrix} \, .
\end{equation}
If we make the ansatz $\Psi_j=\lambda^j$ we obtain in the bulk the equation
\begin{equation}
    \label{bulk}
    t_D \Psi_{j-1}+\mu_k\Psi_j+t_U \Psi_{j+1} = 0 
 \quad \Leftrightarrow \quad p(\lambda)= t_D +\mu_k\lambda +t_U\lambda^2 = 0 \, .
\end{equation}
This quadratic equation will have two solutions $\lambda_\pm$ and the wave function will then have the general form $\Psi_j=A\lambda_-^j+B\lambda_+^j$ where the unknowns $A,B$ are determined by the boundary condition $\mu_k\Psi_0+t_U\Psi_1=0$. In addition, the solution has to be normalizable which means that we must demand $|\lambda_\pm|<1$ generically. I.e., both solutions of the quadratic equation \eqref{bulk} have to lie inside the unit disk in the complex plane. Since $p(z)$ is a polynomial and hence a meromorphic function without poles, $N_{\rm poles}=0$, we can use Cauchy's argument principle
\begin{equation}
    \label{arg_principle}
    N_{\rm in}=\frac{1}{2\pi\im}\oint_{|z|=1} \frac{p'(z)}{p(z)} \, dz 
    =\frac{1}{2\pi\im}\oint_{|z|=1}\frac{2t_U z+\mu_k}{t_U z^2+\mu_k z+t_D}\, dz 
\end{equation}
where $N_{\rm in}=0,1,2$ is the number of roots inside the unit disk. We now want to relate this number to the winding number $I_y$. We note that for a fixed $k\equiv k_x$ we can write the scalar symbol \eqref{HN4} as $F_k(k_y)=\mu_k+t_U\e^{-ik_y}+t_D\e^{ik_y}$, then
\begin{equation}
    \label{slice_winding}
    I_y(k) = \frac{1}{2\pi\im}\int_0^{2\pi} \partial_{k_y}\ln F_k(k_y) dk_y 
    =  -\frac{1}{2\pi\im}\oint_{|z|=1}\frac{F'_k(z)}{F_k(z)} dz
\end{equation}
with $z=\e^{-\im k_y}$. We can use, furthermore, that $F_k(z)=p(z)/z$ which leads to 
\begin{equation}
    \label{slice_winding2}
    I_y(k) = -\frac{1}{2\pi\im}\oint_{|z|=1} \left(\frac{p'(z)}{p(z)}-\frac{1}{z}\right)dz = 1-N_{\rm in} \, .
\end{equation}
This directly shows why the winding number is related to whether or not edge modes do exist. In particular, if $I_y(k)=0$, we have $N_{\rm in}=1$ which means for one solution of the quadratic equation  $|\lambda|<1$ while for the other $|\lambda|>1$. Because a general solution involves both roots, there is no normalizable edge solution in this case. If $I_y(k)=-1$ then $N_{\rm in}=2$ and we have an edge solution which is normalizable and localized at $y=0$. For $I_y(k)=+1$ we have $N_{\rm in}=0$ which means both solutions have $|\lambda|>1$. In this case there is an edge solution of $\mathcal{H}_k^T\mathbf{\Psi}=0$, i.e., there is a localized edge mode if we put the edge at the other side of the tube. 

We have already developed a criterion for the winding number: $I_y\neq 0$ if $E_x\subset E_y$, see Fig.~\ref{Fig_HN}(a). From the explicit solution of the quadratic equation
\begin{equation}
    \label{sol}
    \lambda_\pm =\frac{-\mu_k \pm\sqrt{\mu_k^2-4t_Ut_D}}{2t_U}
\end{equation}
we obtain $\lambda_+\lambda_-=t_D/t_U$. This implies that for both roots to be inside the unit disk we have the necessary condition $|t_D/t_U|<1$. This means that if $E_x\subset E_y$ and $|t_D/t_U|<1$ then $I_y=-1$ while for $E_x\subset E_y$ and $|t_D/t_U|>1$ we have $I_y=+1$ (note that we always assume that the couplings are all real and positive). The same considerations can, of course, also be applied for OBC in x-direction and PBC in y-direction. 

Finally, we can use the boundary condition $\mu_k\Psi_0+t_U\Psi_1=0$ to construct the explicit edge state which is given by 
\begin{equation}
    \Psi_j(k)=\frac{\mu_k/t_U+\lambda_+}{\lambda_+-\lambda_-}\lambda_-^j-\frac{\mu_k/t_U+\lambda_-}{\lambda_+-\lambda_-}\lambda_+^j
\end{equation}
for each allowed $k$ value and provided that $|\lambda_\pm|<1$, which, as we have already shown, is equivalent to having winding $I_y=-1$. In a finite system there are $N_x$ allowed $k$ values and thus $N_x$ edge states. In the limit $N_x\to\infty$ this means, in particular, that we will have a continuous family of edge states with a constant index density $I_y$.

\subsection{Open boundaries in both directions}
So far we have considered the case of a tube, i.e., PBC in one direction and OBC in the other direction. The other interesting question is what happens if we consider OBC in both directions. We have already shown that $(I_x,I_y)=(\pm 1,0)$ or $(0,\pm 1)$, which means that it is impossible to have modes which are localized along the x and y directions at the same time. This excludes, in particular, the existence of corner modes, see Sec.~\ref{Quarter_gapless}, and we expect instead to have the same protected family of edge modes as in the tube case discussed in the previous subsection.  

In Eq.~\eqref{HN2}, we have chosen the couplings in the x-direction to form the inner block $h_0=t_L S_{N_x}+t_R S^T_{N_x}$ of the BTTB matrix. Note that by exchanging the pairs of couplings $(t_R,t_L)\leftrightarrow(t_U,t_D)$ we can also equivalently make the couplings in y-direction form the inner block. For OBC in both directions the choice as in Eq.~\eqref{HN2} is convenient to discuss edge modes in the y-direction while the other choice is convenient to discuss edge modes in the x-direction. We note that a winding number defined by the block symbol $h(k_y)$, see Eq.~\eqref{HN_winding}, is well defined irrespective of whether the $N_x\times N_x$ blocks $h_j$ have open or periodic boundary conditions. With the choice as in \eqref{HN2} we can diagonalize the matrix $h_0$ by performing first a similarity transform $\tilde h_0=D^{-1}h_0D$ with $D=\mbox{diag}(1,\sqrt{t_R/t_L},t_R/t_L,\cdots)$ followed by a diagonalization $h_0^d=U^T\tilde h_0U$ with
\begin{equation}
    \label{OBC1}
   U_{ij}=\sqrt{\frac{2}{N_x+1}}\sin\left(\frac{\pi}{N_x+1}ij\right) \, .
\end{equation}
The eigenvalues are then given by
\begin{equation}
    \label{OBC2}
    \lambda_n = 2\sqrt{t_R t_L}\cos\left(\frac{\pi}{N_x+1}n\right)
\end{equation}
with $n=1,\cdots,N_x$. Accordingly, we can define a scalar function
\begin{equation}
    \label{OBC3}
   \tilde F(k_x,k_y)=2\sqrt{t_R t_L}\cos k_x + t_U \e^{-ik_y}+t_D \e^{ik_y} 
\end{equation}
allowing us to express the block symbol $h(k_y)$ which determines the winding number as 
\begin{equation}
    \label{OBC4}
    \det h(k_y)=\prod_{n=1}^{N_x} \tilde F(\pi n/(N_x+1),k_y) \, .
\end{equation}
We note that \eqref{OBC3} should not be understood as an alternative bulk symbol which is always given by $F(k_x,k_y)$, see Eq.~\eqref{HN4}. In contrast to the bulk symbol, the ellipse $E_x(k_x)$ is now replaced by $2\sqrt{t_Rt_L}\cos k_x$ which is a real interval $\ell_x= [-2\sqrt{t_Rt_L},2\sqrt{t_Rt_L}]$. We therefore have a well-defined winding number in the open boundary case provided that $\det h(k_y)\neq 0$ for all $k_y$, which geometrically means that the ellipse $E_y(k_y)$ shifted by any $x\in \ell_x$ never crosses the origin. If this is the case, then $\mathcal{I}=N_xI_y$ confirming our expectation that the additional edge does not change the bulk-boundary correspondence as long as this edge remains gapped. The exact same considerations, after exchanging the pairs of couplings $(t_R,t_L)\leftrightarrow(t_U,t_D)$, shows that the winding controlling the number of edge modes in x-direction is $\mathcal{I}=N_yI_x$. Since $I_x$ and $I_y$ can never be non-zero at the same time in this model, the edge modes along the two directions are completely independent of each other which is fully consistent with the general considerations in subsection \ref{Quarter_gapless}.

\section{Extended Hatano-Nelson model}
\label{Sec_HN_ext}
Next, we want to extend the Hatano-Nelson model so that it allows for windings $I_x$ and $I_y$ to be non-zero at the same time. In this case, we might expect that there exist modes which are localized in both directions simultaneously.  

\begin{figure}[!htp]
    \centering
    \includegraphics[width=0.5\linewidth]{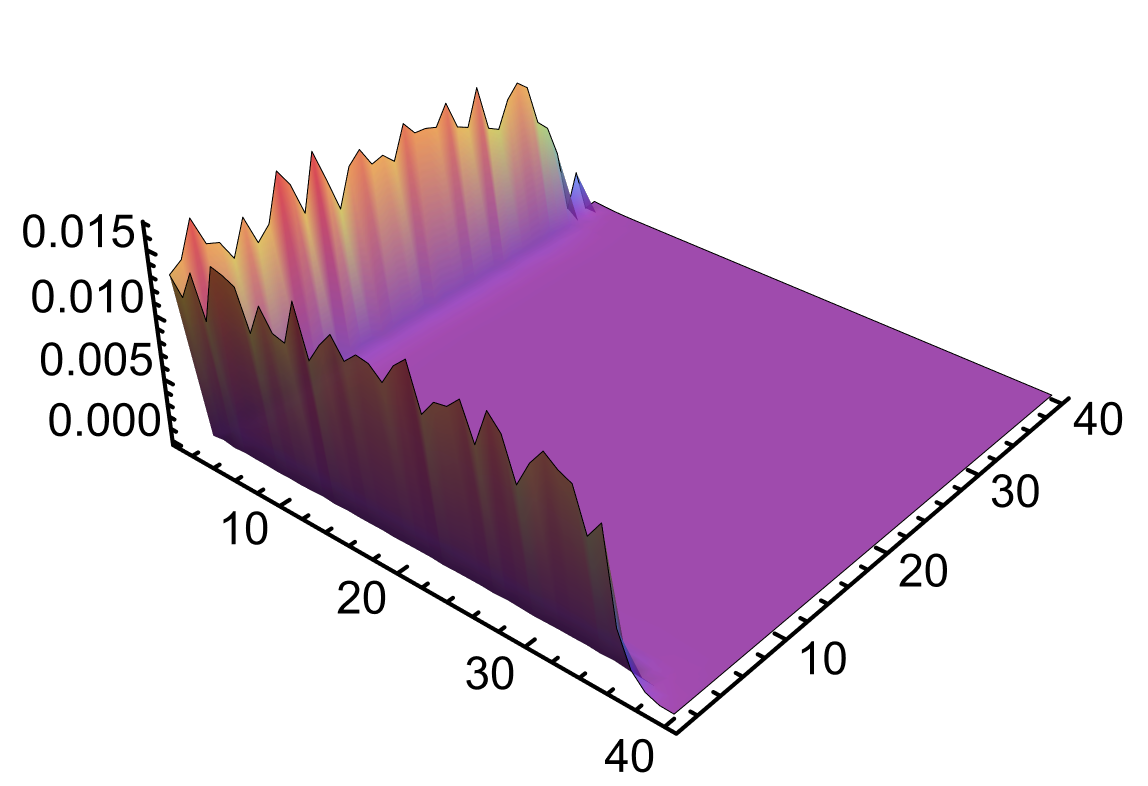}
    \caption{Edge state belonging to the smallest singular value of a $40\times 40$ extended Hatano-Nelson model \eqref{HN_ext} with a perturbation $\varepsilon=0.1$ averaged over $100$ samples showing localization along two edges but no corner mode.}
    \label{Fig_HN_ext}
\end{figure}

One of the simplest extensions of the Hatano-Nelson model that allows for a phase with both slice windings non-zero is obtained by adding diagonal couplings which necessarily tie together the two spatial directions. The Hamiltonian of the Hatano-Nelson model with diagonal couplings is given by
\begin{eqnarray}
    \label{HN_ext}
    H=\sum_{i,j} && \left[ t_R c_{i+1,j}^\dagger c_{i,j}+t_L c_{i,j}^\dagger c_{i+1,j} 
      +t_D c_{i,j+1}^\dagger c_{i,j}+t_U c_{i,j}^\dagger c_{i,j+1} \right.\nonumber \\
    && \left. +t_{NE} c_{i+1,j}^\dagger c_{i,j+1}+t_{SW} c_{i,j+1}^\dagger c_{i+1,j} 
     +t_{NW}c_{i,j}^\dagger c_{i+1,j+1}+t_{SE} c_{i+1,j+1}^\dagger c_{i,j}\right] \, .
    \end{eqnarray}
The block symbol $h(k)$ of this BTTB matrix is then given by $h(k_y)=h_0+h_1\e^{-\im k_y}+h_{-1}\e^{\im k_y}$ with $h_0=t_L S_{N_x}+t_R S_{N_x}^T$, $h_1=t_U\mathbb{I}_{N_x}+t_{NW}S_{N_x}+t_{NE}S_{N_x}^T$, and $h_{-1}=t_D\mathbb{I}_{N_x}+t_{SW}S_{N_x}+t_{SE}S_{N_x}^T$. The second Fourier transform then yields the scalar symbol
\begin{eqnarray}
    \label{HN_ext2}
    F(k_x,k_y) &=& t_L \e^{-\im k_x}+t_R\e^{\im k_x}+t_U\e^{-ik_y}+t_D\e^{\im k_y} \nonumber \\
    &+& t_{NW}\e^{-\im(k_x+k_y)} +t_{NE}\e^{\im(k_x-k_y)} 
    + t_{SW}\e^{-\im(k_x-k_y)} +t_{SE}\e^{\im(k_x+k_y)} \, .
\end{eqnarray}
In Fig.~\ref{Fig_HN_ext} we show the state that belongs to the smallest singular value for a set of parameters such that $(I_x,I_y)=(-1,-1)$.
Clearly, the state corresponds to an edge mode of the type \eqref{edge} and not a corner mode of the type \eqref{corner}. As discussed earlier in Sec.~\ref{Quarter_gapless}, this is expected to be the generic scenario for a scalar symbol. Here the two half-plane problems each support a 1-parameter family of edge-localized solutions. On a finite rectangle these families generically hybridize near the corners, producing states with support along both edges, Eq.~\eqref{edge}. A genuinely corner-localized state \eqref{corner} would require additional constraints ensuring simultaneous exponential decay in both directions (e.g., a factorization, see the example in the next section), and is therefore non-generic. Numerically, we find that the number of singular values corresponding to edge modes in this specific model scales as $N_x +N_y -1$. This is consistent with the scaling expected from the slice windings which predicts that the number of such modes is always $\mathcal{O}(N_x+N_y)$ but the precise number will depend on microscopic details of the hybridization between the two families of edge modes. I.e., the precise count found numerically for the extended Hatano-Nelson model is not necessarily generic. 

\section{Coexisting edge and corner modes}
\label{Sec_cor}
As we discussed in Sec.~\ref{Sec_gen}, there are two types of systems that can show corner modes: (i) systems with families of gapless edge modes where additional restrictions lead to the existence of an additional small number of corner modes, and (ii) systems where both the bulk and the edges are gapped and corner modes are the only type of gapless excitations in the system. We note again that in the first class of systems, corner modes are protected by a gap separating them from the edge modes but not by a topological index. The protection by a true topological corner index is only realized in case (ii). 

To obtain a corner mode in case (i), we need as necessary conditions (a) a phase where both $I_x$ and $I_y$ are non-zero at the same time, and (b) a symbol $F(k_x,k_y)$ that is a matrix and not a scalar. Physically, this means that at a minimum we have to consider a two-band model. To obtain a case (i) example, we want to build a model where the symbol is sublattice symmetric and each block has a product structure between $x$ and $y$. A Hamiltonian that has the required structure is given by 
\begin{eqnarray}
    \label{product}
    H &=& \sum_{i,j} (\gamma_x \gamma_y a_{i,j}^\dagger b_{i,j} + \alpha_x\alpha_y b_{i,j}^\dagger a_{i,j} ) 
    + \sum_{i,j} (\beta_x\alpha_y b_{i,j}^\dagger a_{i+1,j} + \lambda_x\gamma_y a_{i+1,j}^\dagger b_{i,j}) \\
    &+& \sum_{i,j} (\gamma_x\lambda_y a_{i,j+1}^\dagger b_{i,j} + \alpha_x\beta_y b_{i,j}^\dagger a_{i,j+1} ) 
    + \sum_{i,j} (\lambda_x\lambda_y a_{i+1,j+1}^\dagger b_{i,j} + \beta_x\beta_y b_{i,j}^\dagger a_{i+1,j+1}). \nonumber
\end{eqnarray}
The symbol is then a $2\times 2$ matrix
\begin{equation}
    \label{product2}
    F(k_x,k_y) = \begin{pmatrix}
        0 & q_L(k_x,k_y) \\
        q_R(k_x,k_y) & 0
    \end{pmatrix}
\end{equation}
with 
\begin{equation}
    \label{product3}
    q_L(k_x,k_y) = (\gamma_x +\lambda_x \text{e}^{ik_x})(\gamma_y+\lambda_y\text{e}^{ik_y}) \, , \qquad
    q_R(k_x,k_y) = (\alpha_x +\beta_x \text{e}^{-ik_x})(\alpha_y+\beta_y\text{e}^{-ik_y}) \, .  
\end{equation}
The example is set up in a way that both $q_L$ and $q_R$ have product structure. We will see below that it is this product structure that guarantees the existence of a corner mode. For a system with $N_x=N_y=2$ unit cells and an $(a,b)$ basis along x, the sub-matrices are explicitly
\begin{eqnarray}
    \label{product4}
    h_0 &=& \begin{pmatrix}
        0 & \gamma_x\gamma_y & 0 & 0 \\
        \alpha_x\alpha_y & 0 & \beta_x\alpha_y & 0 \\
        0 & \lambda_x\gamma_y & 0 & \gamma_x\gamma_y \\
        0 & 0 & \alpha_x\alpha_y & 0
    \end{pmatrix} \\
    h_1 &=& \begin{pmatrix}
        0 & 0 & 0 & 0 \\
        \alpha_x\beta_y & 0 & \beta_x\beta_y & 0 \\
        0 & 0 & 0 & 0 \\
        0 & 0 & \alpha_x\beta_y & 0
    \end{pmatrix} \, , \qquad
    h_{-1} = \begin{pmatrix}
        0 & \gamma_x\lambda_y & 0 & 0 \\
        0 & 0 & 0 & 0 \\
        0 & \lambda_x\lambda_y & 0 & \gamma_x\lambda_y \\
        0 & 0 & 0 & 0
    \end{pmatrix} \nonumber
\end{eqnarray}
with all other sub-matrices equal to zero.

We are trying to construct a corner mode which lives at the top left corner. To the right and to the bottom the system is infinite. Then the equations are
\begin{equation}
    \label{product5}
    \gamma_x\gamma_y \Psi^B_{1,1} = 0,\quad
    \lambda_x\gamma_y \Psi^B_{1,1} + \gamma_x\gamma_y \Psi^B_{2,2} = 0,\quad \cdots
\end{equation}
which recursively implies that $\Psi^B\equiv 0$ if the couplings are non-zero. The mode lives on the A sublattice only. For the A sublattice, we get always the same bulk equation without any additional boundary equations
\begin{equation}
    \label{product6}
\alpha_x\alpha_y\Psi^A_{m,n}+\beta_x\alpha_y\Psi^A_{m+1,n}+\alpha_x\beta_y\Psi^A_{m,n+1}+\beta_x\beta_y\Psi^A_{m+1,n+1}=0
\end{equation}
We can make the ansatz $\Psi^A_{m,n}=w^m z^n$. Then \eqref{product6} becomes
\begin{equation}
    \label{product7}
 \alpha_x\alpha_y +  \beta_x\alpha_y w + \alpha_x\beta_y z + \beta_x\beta_y zw =  0 \;\Leftrightarrow\;
 (\alpha_x+\beta_x w)(\alpha_y+\beta_y z) = 0
\end{equation}
which is identical to the condition $q_R(w,z)=0$, see Eq.~\eqref{product3}, with $w=\e^{-\im k_x}$ and $z=\e^{-\im k_y}$. This condition is fulfilled by setting either $w=w_0=-\alpha_x/\beta_x$ or $z=z_0=-\alpha_y/\beta_y$ and if $|w|<1$ or $|z|<1$, this is an exponentially decaying solution. However, the other parameter is free so that the kernel for the infinite quarter plane would be infinite dimensional. This is a result of the exact product structure. However, if we consider a finite quarter plane then we have additional boundary conditions coming from the right boundary and the bottom boundary of the plane which give the two conditions
\begin{equation}
    \label{product9}
    \alpha_x(\alpha_y+\beta_y z) = 0 \, , \qquad
    (\alpha_x+\beta_x w)\alpha_y = 0
\end{equation}
which must be fulfilled simultaneously. This implies that $(w,z)=(w_0,z_0)$ is the only solution in this case if both $\alpha_{x,y}$ are non-zero. This solution is an exponentially decaying corner mode if 
\begin{equation}
    \label{product8}
|\alpha_x| < |\beta_x| \quad \mbox{and} \quad |\alpha_y|<|\beta_y|
\end{equation}
which are exactly the conditions for both windings $I_R^x$ and $I_R^y$ to be non-zero, see Eq.~\eqref{product3}. In addition to this corner mode, there will also still be edge modes. The exact number of these modes will depend on all four winding numbers $I_L^x,I_L^y,I_R^x,I_R^y$. While we do need $|I_R^x|=|I_R^y|=1$ to have a corner mode, $I_L^x$ and $I_L^y$ only influence the number of edge modes. While the slice windings directly predict the minimal number of protected edge modes for a scalar symbol if we have periodic boundary conditions in one direction, this is different here where we have a matrix-valued symbol and open boundary conditions in both directions. First of all, the index for a matrix-valued symbol only predicts the difference between the number of left and right zero modes similar to the Hermitian Chern insulator. In addition, if we have open boundary conditions in both directions, edge modes can hybridize which can also affect their numbers. So the only thing we can say in the general case is that the number of edge modes will scale as $\mathcal{O}(N_x+N_y)$ with increasing system size but the slice windings alone do not fix their exact number. 
\begin{figure}[htp]
    \centering
    \includegraphics[width=0.48\columnwidth]{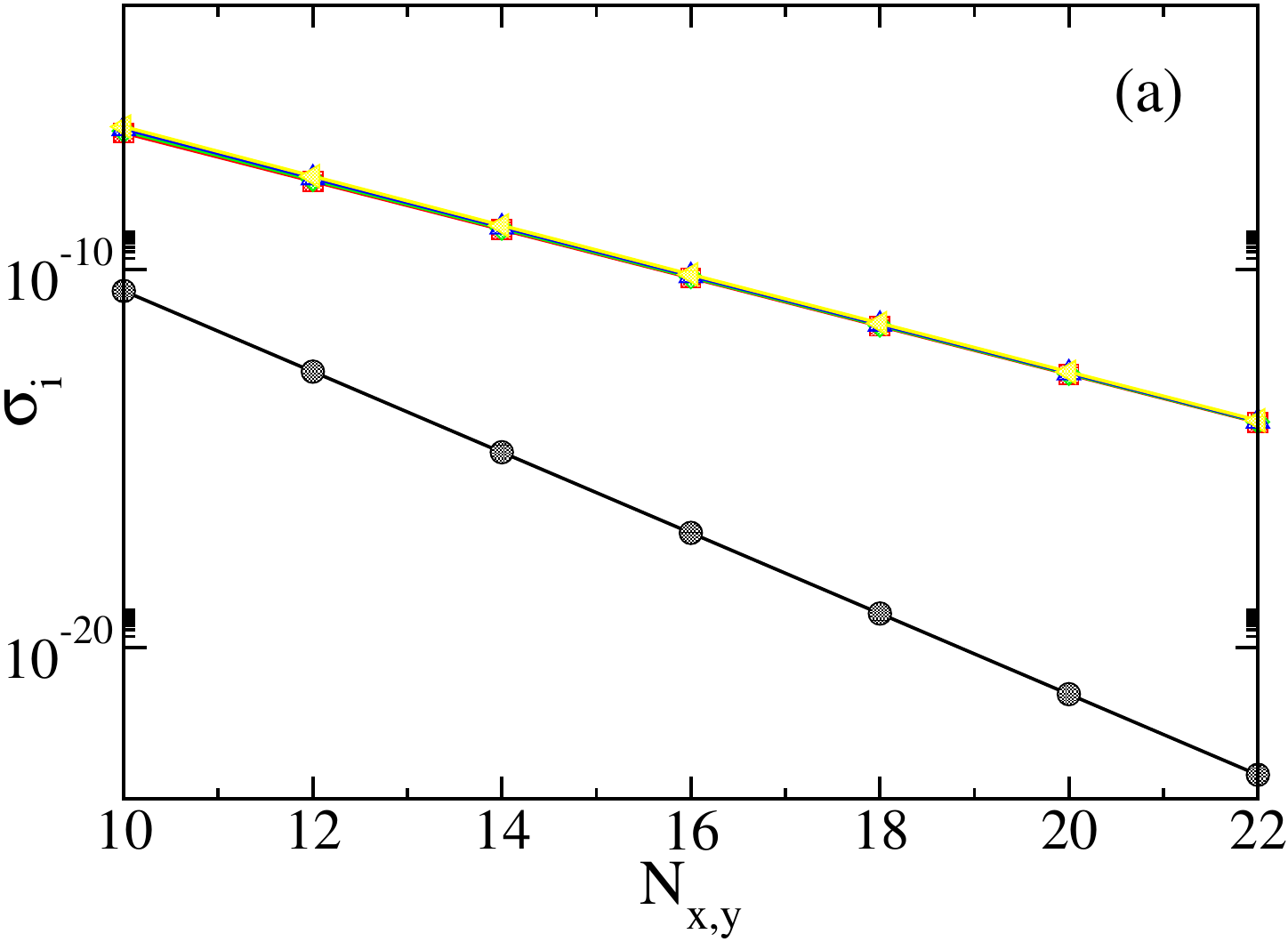}
    \includegraphics[width=0.48\columnwidth]{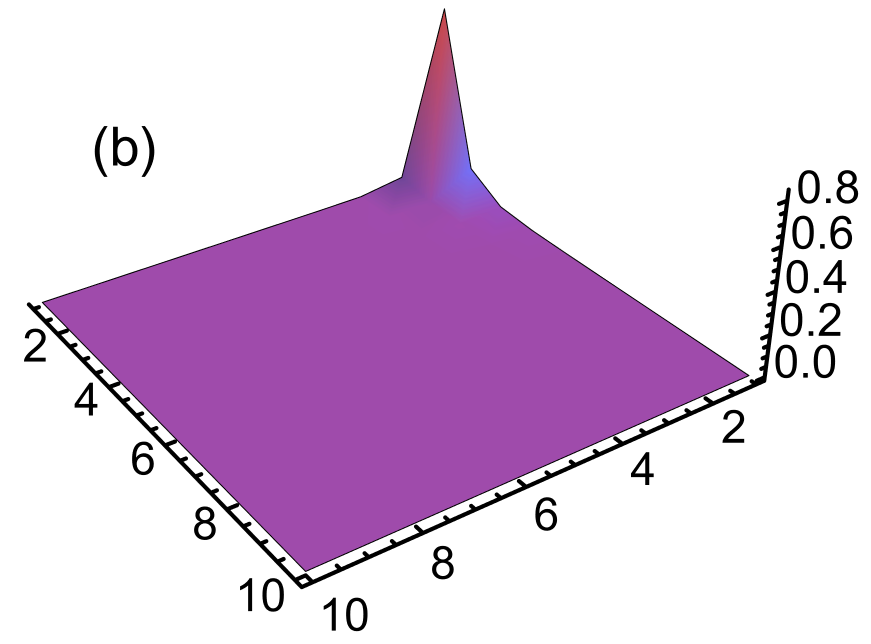}
    \caption{(a) The five smallest singular values for the Hamiltonian \eqref{product} as function of unit cells in the phase with $(I_L^x,I_L^y,I_R^x,I_R^y)=(0,0,-1,-1)$. The singular value belonging to the corner mode scales much faster to zero than those of the edge modes (almost on top of each other on this scale). (b) Spectrally protected corner mode of the Hamiltonian \eqref{product} in the phase with $(I_L^x,I_L^y,I_R^x,I_R^y)=(0,0,-1,-1)$ obtained after averaging over $10$ samples with perturbation strength $\varepsilon=0.1$.}
    \label{Fig1_product}
\end{figure}
The scaling of the smallest five singular values for the Hamiltonian \eqref{product} with $N_x\times N_y$ unit cells and $N_x=N_y$ is shown in Fig.~\ref{Fig1_product}(a) for a phase with windings $(0,0,-1,-1)$. Note that the system sizes here are smaller than in the previous two examples because the unit cell now consists of two sites and, furthermore, accurately resolving the smallest singular value and the corresponding singular vector requires multi-precision numerics. For this specific case, we find numerically that there are $N_x+N_y-1$ singular values which scale to zero. The smallest one of them, which belongs to the spectrally protected corner mode shown in Fig.~\ref{Fig1_product}(b), decays faster than those belonging to protected edge modes. This behavior can be understood as follows: In a finite quarter-plane geometry, boundary modes of the infinite Toeplitz operator are no longer exact kernel vectors because the additional boundaries violate the bulk recursion relations. The resulting residual norm $\|H\Psi\|\sim\e^{-N}$ is generated entirely at these distant boundaries, where the wave function amplitude is exponentially small. For an edge mode, the wave function decays only in one direction and therefore has comparable exponential weight along an entire boundary segment of length $\mathcal{O}(N_x)$ or $\mathcal{O}(N_y)$. The residual norm thus receives contributions from a macroscopic number of boundary sites of the same exponential order. For a corner mode, by contrast, the wave function decays exponentially in both spatial directions. Along the distant boundaries, its amplitude continues to decay along the boundary itself, so that the total boundary contribution to the residual norm is exponentially summable.

\section{Topologically protected corner modes}
\label{Sec_BBH}
The best-known example of a model showing protected corner modes is the Benalcazar-Bernevig-Hughes (BBH) model \cite{Benalcazar2017Science}, an example of a Hermitian higher-order topological insulator. If we take a quarter plane, then, importantly, not only the bulk needs to be gapped but also the edges in order to allow for topologically protected corner modes. In Ref.~\cite{Benalcazar2017Science}, the topological character of the model is discussed in terms of a quantized quadrupole moment. This requires certain crystalline symmetries such as bulk inversion symmetry and reflection symmetries to be present. However, a stable topological index and corresponding corner modes do exist even if all crystalline symmetries are broken as long as chiral symmetry is preserved and the bulk and edges remain gapped \cite{Hayashi2019Toeplitz}.

Here we want to further generalize the BBH model and show that we can also break Hermiticity by introducing non-reciprocal hoppings while still having a topological phase with protected corner modes. The BBH model is a four-band model and, as in Ref.~\cite{Benalcazar2017Science}, we define the following matrices acting on the degrees of freedom within the unit cell: $\tilde\Gamma_0=\sigma_3\otimes\mathbb{I}$, $\tilde\Gamma_k=\sigma_2\otimes\sigma_k$ with $k=1,2,3$, and $\tilde\Gamma_4=\sigma_1\otimes\mathbb{I}$. Here, $\sigma_k$ are the three Pauli matrices and $\mathbb{I}$ is the $2\times 2$ identity. The symbol of a generalized BBH-Hamiltonian with non-reciprocal couplings can then be written as
\begin{eqnarray}
    \label{BBH1}
    \tilde h(k_x,k_y)\!\! &=& \!\!\gamma_{x1}\tilde\Gamma_4 + \gamma_{x2}\tilde\Gamma_3 + \gamma_{y1}\tilde\Gamma_2 +\gamma_{y2}\tilde\Gamma_1 
    +\frac{1}{2}(\lambda_{x1}^R \tilde\Gamma_4-\im \lambda_{x2}^R\tilde\Gamma_3)\e^{\im k_x}+\frac{1}{2}(\lambda_{x1}^L \tilde\Gamma_4+\im \lambda_{x2}^L\tilde\Gamma_3)\e^{-\im k_x} \nonumber \\
    &+&\frac{1}{2}(\lambda_{y1}^R \tilde\Gamma_2-\im \lambda_{y2}^R\tilde\Gamma_1)\e^{\im k_y}+\frac{1}{2}(\lambda_{y1}^L \tilde\Gamma_2+\im \lambda_{y2}^L\tilde\Gamma_1)\e^{-\im k_y} 
\end{eqnarray}
and the standard Hermitian BBH model with crystalline symmetries as written in Eq.~(6) of Ref.~\cite{Benalcazar2017Science} is recovered if "$R=L$" (reciprocal couplings), $\gamma_{x2}=\gamma_{y2}=0$, $\gamma_{x1}=\gamma_{y1}$, and $\lambda_{x1}=\lambda_{y1}=\lambda_{x2}=\lambda_{y2}$. The important step in understanding why this model does have a stable corner index even without crystalline symmetries is realizing that, under the unitary transformation $U$ defined by 
\begin{equation}
    \label{BBH2}
    U =\begin{pmatrix}
0 & -1 & 0 & 0 \\
0 & 0 & 1 & 0\\
0 & 0 & 0 & -1\\
1 & 0 & 0 & 0
    \end{pmatrix} \, ,
\end{equation}
the Hamiltonian \eqref{BBH1} can be brought into an equivalent form showing that it is constructed from one-dimensional sublattice-symmetric Hamiltonians along the x and y directions. Under the transformation $U^T \tilde\Gamma_j U=\Gamma_j$ we obtain new matrices $\Gamma_j$ acting in the internal orbital space which are given by $\Gamma_0=-\sigma_3\otimes\sigma_3$, $\Gamma_1=\sigma_3\otimes\sigma_2$, $\Gamma_2=\sigma_3\otimes\sigma_1$, $\Gamma_3=\sigma_2\otimes\mathbb{I}$, and $\Gamma_4=\sigma_1\otimes\mathbb{I}$. By replacing $\tilde\Gamma_j\to\Gamma_j$ it can be seen from Eq.~\eqref{BBH1} that the transformed symbol $h(k_x,k_y)=U^T \tilde h(k_x,k_y) U$ can be written as \cite{Hayashi2019Toeplitz}
\begin{equation}
    \label{BBH3}
    h(k_x,k_y) = h_x(k_x)\otimes\mathbb{I} + \sigma_3\otimes h_y(k_y)
\end{equation}
with
\begin{eqnarray}
    \label{BBH4}
    h_x(k_x) &=& \gamma_{x1}\sigma_1 +\gamma_{x2}\sigma_2 + \frac{1}{2}(\lambda_{x1}^R\sigma_1-\im\lambda_{x2}^R\sigma_2)\e^{\im k_x} 
    + \frac{1}{2}(\lambda_{x1}^L\sigma_1+\im\lambda_{x2}^L\sigma_2)\e^{-\im k_x} \nonumber \\
    h_y(k_y) &=& \gamma_{y1}\sigma_1 +\gamma_{y2}\sigma_2 + \frac{1}{2}(\lambda_{y1}^R\sigma_1-\im\lambda_{y2}^R\sigma_2)\e^{\im k_y} 
    + \frac{1}{2}(\lambda_{y1}^L\sigma_1+\im\lambda_{y2}^L\sigma_2)\e^{-\im k_y} \, .\nonumber
\end{eqnarray}
Note that this system still has sublattice symmetry with $\{\Pi,h(k_x,k_y)\}=0$ where $\Pi=\sigma_3\otimes\sigma_3$. Eq.~\eqref{BBH4} represents two sublattice symmetric one-dimensional systems which describe the edges of the quarter plane. A corner state in this factorizable model is then of the form $\Psi_{\rm{corner}}\sim\Psi_x\otimes\Psi_y$ if both $\Psi_x$ and $\Psi_y$ are edge localized in one dimension and share a corner. More precisely, we enumerate the unit cells starting from the top left corner of the finite rectangle with $N_x\times N_y$ unit cells. We call the edge at $x=1$ or $y=1$ of the one-dimensional chains \eqref{BBH4} a left corner and the one at $x=N_x$ or $y=N_y$ a right corner. If we consider, for example, the $(x,y)=(1,1)$ corner then we have $K_{11}=n_x^L n_y^L$ corner modes localized at this corner if $n_x^L$ is the number of edge modes localized at the left edge in the $h_x$ chain and $n_y^L$ the number of left-localized modes in the $h_y$ chain. This means that the total number of topologically protected corner modes at all four corners of the finite system can be expressed as 
\begin{equation}
    \label{BBH5}
    K=(n_x^L + n_x^R)(n_y^L + n_y^R) \, .
\end{equation}
This counting reflects the fact that a corner mode exists if and only if an edge-localized mode of $h_x$ and an edge-localized mode of $h_y$ meet at the same corner. $K$ is also the number of topologically protected singular values that go to zero with increasing system size. Next, we need to connect the number of edge localized states with the topologies of the sublattice symmetric chains. Both symbols $h_x(k_x)$ and $h_y(k_y)$ are $2\times 2$ matrices with the diagonal elements being zero reflecting the sublattice symmetry. For each of the off-diagonal elements, we can define a winding number using Eq.~\eqref{winding} if the determinant of this block is always non-zero, which means that it has a point gap. I.e., the two-dimensional symbol $h(k_x,k_y)$ constructed from the one-dimensional symbols, see Eq.~\eqref{BBH3}, is topologically characterized by the four winding numbers $(I^x_1,I^x_2,I^y_1,I^y_2)$. For this specific model, the possible winding numbers are $I^{x,y}_j\in\{-1,0,1\}$. A winding of $+1$ is associated with a right-localized mode and a winding of $-1$ with a left-localized mode. We therefore obtain
\begin{equation}
    \label{BBH6}
    n_x^R = \max(I^x_1,0) + \max(I^x_2,0) \, , \qquad
    n_x^L = \max(-I^x_1,0) + \max(-I^x_2,0)
\end{equation}
and similarly for $n_y^{R,L}$. Note that for this one-dimensional two-band sublattice symmetric model, each off-diagonal block contributes at most one edge-localized mode, so the absolute values of the winding numbers directly count the number of left- or right-localized states.

As a concrete example, we consider the case with windings $(I^x_1,I^x_2,I^y_1,I^y_2)=(1,-1,1,-1)$. In this case, we expect $K=4$ singular values which go to zero with increasing system size. They correspond to four corner modes, each one of them localized at a different corner of the system with $N_x\times N_y$ unit cells.
\begin{figure}
    \centering
    \includegraphics[width=0.5\columnwidth]{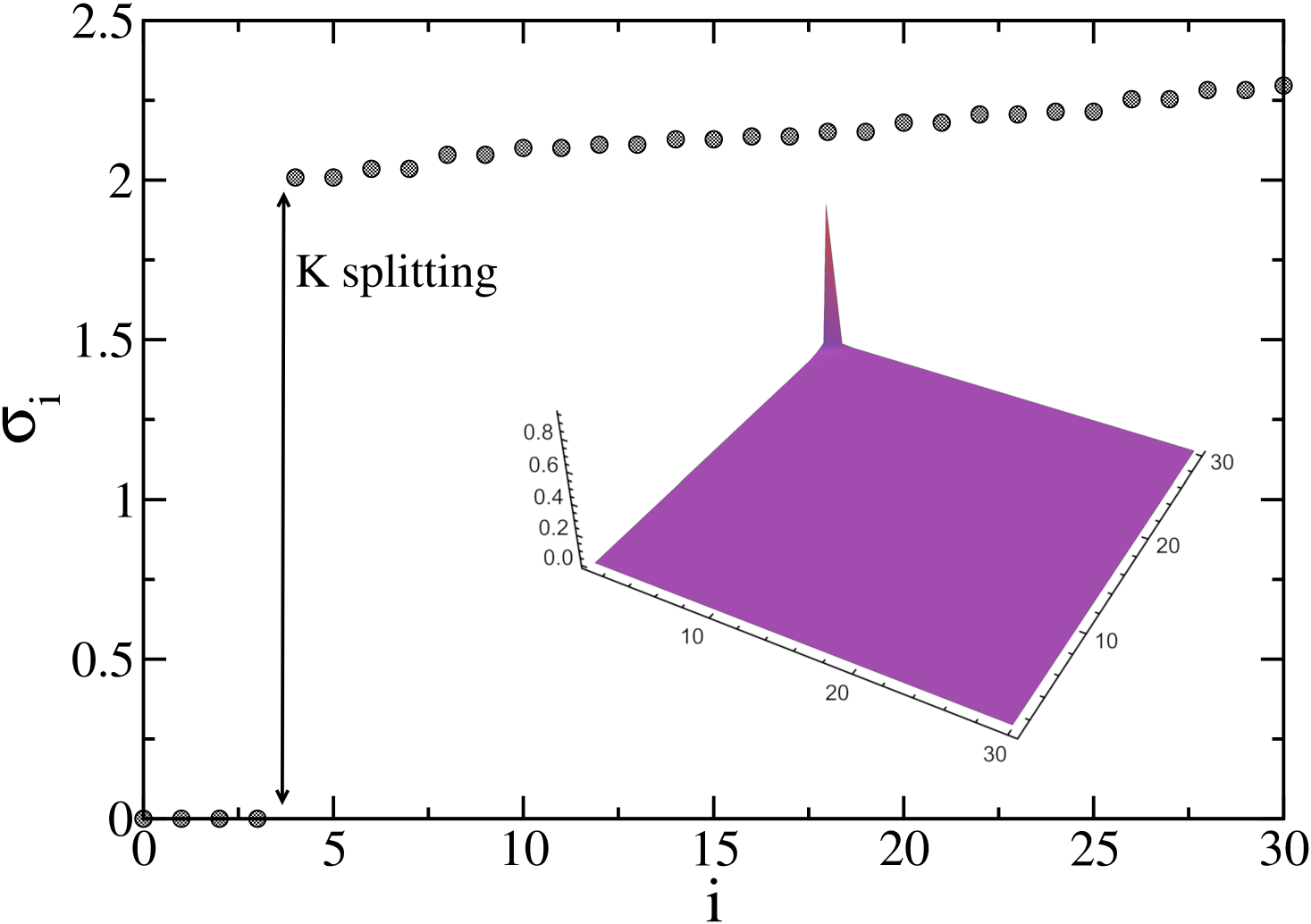}
    \caption{Smallest singular values---averaged over $20$ samples with a disorder $\varepsilon=0.1$ that respects the sublattice symmetry---for the non-Hermitian BBH model \eqref{BBH3} on a lattice with $30\times 30$ unit cells in the topological $(1,-1,1,-1)$ phase where both the bulk and the edges are gapped. The inset shows one of the four corner modes. The parameters are $\gamma_{x1}=0.2,\gamma_{x2}=0.1,\gamma_{y1}=0.5,\gamma_{y2}=0.05,\lambda_{x1}^R=\lambda_{x2}^R=\lambda_{y1}^R=\lambda_{y2}^R=2.7,\lambda_{x1}^L=2.5,\lambda_{x2}^L=2.6,\lambda_{y1}^L=2.4,\lambda_{y2}^L=2.8$ and all symmetries except for the sublattice symmetry $\Pi$ are broken.}
    \label{Fig_BBH}
\end{figure}
These expectations are confirmed numerically, see Fig.~\ref{Fig_BBH}, where the K-splitting between four extremely small singular values corresponding to the corner modes and the rest of the large bulk singular values is clearly visible. In the inset, one of the four corner modes is shown.  

To summarize, one can generalize the BBH model to a non-Hermitian model by allowing for non-reciprocal hoppings. Crucially, one can write this model as a sum of tensor products involving one-dimensional sublattice symmetric models and the corner index can be written in terms of the winding numbers of these chains. Since the winding numbers are invariant under point-gap-preserving deformations, the corner modes remain stable under generic non-Hermitian perturbations that preserve sublattice symmetry and the bulk and edge gaps. In the Hermitian limit of the standard BBH model, the four winding numbers reduce to the two BBH edge polarizations, and Eq.~\eqref{BBH5} reproduces the usual quadrupole-phase corner counting.

\section{Conclusions}
\label{Sec_Con}
In this work, we have established bulk-boundary correspondences for two-dimensional non-Hermitian Gaussian models. One of the main messages of this paper is that eigenvalue spectra of non-normal operators are generically unstable under changes of boundary conditions and disorder and therefore cannot reflect stable topological properties of a physical system. On the other hand, the singular value spectrum is stable and admits a precise bulk-boundary correspondence via index and K-splitting theorems.

More specifically, the bulk Hamiltonians define infinite Toeplitz operators which can be truncated either to half or quarter planes. Well-established index theorems and K-splitting theorems in Toeplitz theory then connect the bulk topology, described via a non-trivial index, with singular values in the truncated operator which go to zero with increasing system size and are separated from the bulk singular value spectrum by a gap. Crucially, topologically protected boundary modes are only exact eigenstates in the thermodynamic limit. For a finite system, they correspond in general to vectors which only get mapped exponentially close to zero by the Hamiltonian. We dubbed these modes 'hidden zero modes' because they cannot be found in an eigen-decomposition. Physically, these modes are relevant because they correspond to extremely long-lived excitations, which are essentially indistinguishable from true eigenmodes for large systems. We also stressed that half and quarter plane truncations are inequivalent and require distinct index-theoretic considerations which lead to different boundary phenomena.

We derived a number of general results for these two different types of truncations and for different classes of symbols. We showed that if the symbol of the Hamiltonian is scalar, then the topology of the system is fully characterized by the two slice windings $(I_x,I_y)$ and in a system with periodic boundary conditions in one and open boundary conditions in the other, the number of edge states is exactly $N_\alpha |I_\alpha|$ with $\alpha=x,y$ and the side on which the states localize is determined by the sign of $I_\alpha$. Matrix-valued symbols, on the other hand, are different. Here, the bulk index fixes only the difference between the number of left and right zero modes, not their absolute number. This mirrors the case of the Hermitian Chern insulator. Consequently, the slice windings only predict lower bounds for the number of edge modes for matrix-valued symbols. This is not a shortcoming of the theory but rather a mathematical consequence of the fact that in this case the kernels of the infinite Toeplitz operator and of its adjoint can be non-zero at the same time. Irrespective of whether the symbol is scalar or matrix-valued, the slice windings are the only bulk topological invariants in the half plane case for point-gapped systems and fully determine its topological properties. In the quarter plane geometry, on the other hand, there are two cases that need to be distinguished: gapless and gapped edges. In the former case, the slice windings are still the proper topological invariants and, if non-zero, predict families of edge modes. If there is a phase where both slice windings are non-zero at the same time, then families of edge states do exist along both edges and they hybridize at the shared corner. This generically leads to edge modes which extend over both edges but, if additional restrictions are present, can also lead to corner modes which coexist with the edge modes. Importantly, this type of corner mode is only spectrally protected instead of being protected by an additional index theorem. If both the bulk and the edges are gapped, then a proper index can be defined for the quarter plane because in this case the semi-infinite quarter plane is a Fredholm operator with a finite kernel and co-kernel. This index can then again be connected to the properties of the singular value spectrum via a K-splitting theorem. 

We illustrated our general results using four distinct examples. The first and simplest one was a Hatano-Nelson type model on the square lattice which has a scalar symbol. In this model, the only topological phases are characterized by a single slice winding being non-zero, leading to the existence of a family of edge states but excluding the existence of corner modes. We also showed explicitly that any eigenvalue-based bulk-boundary correspondence fails because a breaking of translational invariance---which is unrelated to topology---leads to a pseudospectrum which fills the entire image of the symbol $F(k_x,k_y)$. In particular, because the bulk is gapped in a topological phase, $F(k_x,k_y)\neq 0$ for all $k_x,k_y$, there are no eigenvalues in the pseudospectrum which are close to zero. 

In the second example, we generalized the Hatano-Nelson model by also allowing diagonal couplings. This led to topological phases where both slice windings are non-zero. However, the result was edge states localized along both edges, not corner modes, highlighting that in the case of gapless edges $I_x\neq 0$ and $I_y\neq 0$ is a necessary but not sufficient condition for corner modes to exist. 

In the third example, we introduced a Hamiltonian with a matrix-valued symbol which has a product structure. We showed that this model also has phases with $I_x\neq 0$ and $I_y\neq 0$ and that the additional algebraic structure does lead to the stabilization of corner modes which coexist with families of edge modes. Importantly, these corner modes are not protected by a separate bulk index---the slice windings $I_{x,y}$ remain the only topological invariants---but rather by a hierarchy in the singular value spectrum. The bulk singular values are separated from the boundary singular values by a gap as predicted by the standard K-splitting theorem. Crucially, inside the boundary singular values there is another gap which is caused by the singular values belonging to corner modes decaying parametrically faster with system size than those belonging to edge modes. We explained that this is due to the different residual mismatch of these modes at the finite-size boundaries where these boundary modes are exponentially small.

Our final example was a non-Hermitian generalization of the Benalcazar-Bernevig-Hughes model. We first showed that this model, including the Hermitian limit, is unitarily equivalent to a two-dimensional model generated from one-dimensional chains with sublattice symmetry. We furthermore showed that this model has a genuine quarter plane index which can be calculated from the winding numbers of the sublattice-symmetric chains. This Toeplitz corner index exists because not only the bulk but also the edges are gapped. The K-splitting theorem then directly predicts the number of corner modes in this non-Hermitian system with higher-order topology. Crucially, crystalline symmetries, which were thought to be essential in earlier studies of Hermitian higher-order topology, are not required and we demonstrated this explicitly by breaking all of them. It is fully sufficient that sublattice symmetry is preserved and that the blocks of the sublattice-symmetric chains have a point gap. 

Our results show that proper bulk-boundary correspondences in non-Hermitian systems are operator-theoretic, not spectral. Toeplitz operator theory and K-splitting theorems for the singular values provide the natural framework to systematically study the topological properties of non-Hermitian phases. In addition, this framework also naturally unifies first-order and higher-order topology without the need to invoke crystalline symmetries. An important practical consequence is that singular values provide a reliable finite-size diagnostic of topology in non-Hermitian systems, whereas eigenvalues do not converge uniformly and therefore cannot be used to detect topological boundary modes in finite samples. 

\section*{Acknowledgements}
The author acknowledges support by the National Science and Engineering Research Council (NSERC) of Canada through the Discovery Grants program. The author also gratefully acknowledges the hospitality of Utrecht University and the RPTU Kaiserslautern-Landau where part of this work was performed.

%\bibliography{Literatur}

\end{document}